\documentclass{aastex}

\usepackage[onecolumn,numberedappendix]{emulateapj5}

\newcommand{\gtsima}{$\; \buildrel > \over \sim \;$}
\newcommand{\ltsima}{$\; \buildrel < \over \sim \;$}
\newcommand{\simgt}{\lower.5ex\hbox{\gtsima}}
\newcommand{\simlt}{\lower.5ex\hbox{\ltsima}}
\newcommand{\himpc}{{\hbox {$h^{-1}$}{\rm Mpc}} }
\newcommand{\bfk}{{\mbox{\boldmath $k$}}}
\newcommand{\bfx}{{\mbox{\boldmath $x$}}}
\newcommand{\bfv}{{\mbox{\boldmath $v$}}}
\newcommand{\bfn}{{\mbox{\boldmath $n$}}}
\newcommand{\sbx}{{\mbox{\scriptsize\boldmath $x$}}}
\newcommand{\sbk}{{\mbox{\scriptsize\boldmath $k$}}}
\newcommand{\sbom}{{\mbox{\scriptsize\boldmath $\omega$}}}
\newcommand{\sbth}{{\mbox{\scriptsize\boldmath $\theta$}}}
\newcommand{\bfth}{{\mbox{\boldmath $\theta$}}}
\newcommand{\bfom}{{\mbox{\boldmath $\omega$}}}
\newcommand{\bfeta}{{\mbox{\boldmath $\eta$}}}
\newcommand{\bfJ}{{\mbox{\boldmath $J$}}}
\newcommand{\bfA}{{\mbox{\boldmath $A$}}}
\newcommand{\bfM}{{\mbox{\boldmath $M$}}}
\newcommand{\hMt}{\widehat{M}^{(3)}}
\newcommand{\tnu}{{\widetilde{\nu}}}

\newcommand{\thf}{{\theta_{\rm f}}}
\newcommand{\PL}{P_{\rm LIN}}
\newcommand{\nspec}{{n_{\rm s}}}
\newcommand{\logd}{g_f}

\shortauthors{MATSUBARA} 
\shorttitle{STATISTICS OF SMOOTHED COSMIC FIELDS}

\begin{document}


\title{STATISTICS OF SMOOTHED COSMIC FIELDS IN PERTURBATION THEORY I:
\\ Formulation and useful formulas in second-order perturbation
theory}

\author{Takahiko Matsubara}
\affil{Department of Physics and Astrophysics, Nagoya University,
Chikusa, Nagoya 464-8603, JAPAN
}
\email{taka@a.phys.nagoya-u.ac.jp}

\begin{abstract}

We formulate a general method for perturbative evaluations of
statistics of smoothed cosmic fields, and provide useful formulas in
application of the perturbation theory to various statistics. This
formalism is an extensive generalization of the method used by
\citet{mat94} who derived a weakly nonlinear formula of the genus
statistic in a 3D density field. After describing the general method,
we apply the formalism to a series of statistics, including genus
statistics, level-crossing statistics, Minkowski functionals, and a
density extrema statistic, regardless of the dimensions in which each
statistic is defined. The relation between the Minkowski functionals
and other geometrical statistics is clarified. These statistics can be
applied to several cosmic fields, including 3D density field, 3D
velocity field, 2D projected density field, and so forth. The results
are detailed for second order theory of the formalism. The effect of
the bias is discussed. The statistics of smoothed cosmic fields as
functions of rescaled threshold by volume-fraction are discussed in
the framework of second-order perturbation theory. In CDM-like models,
their functional deviations from linear predictions plotted against
the rescaled threshold are generally much smaller than that plotted
against the direct threshold. There is still slight meat-ball shift
against rescaled threshold, which is characterized by asymmetry in
depths of troughs in the genus curve. A theory-motivated asymmetry
factor in genus curve is proposed.

\end{abstract}



\setcounter{equation}{0}
\section{INTRODUCTION}
\label{sec1}

Any conceivable theory of structure formation in the universe predicts
statistical properties of observable quantities. Therefore, the
analysis of the present inhomogeneity of the universe is inevitably
statistical. However, what kind of statistics is useful is not
obvious, since we can invent infinite number of statistics to be
analyzed. Whether we can adopt better statistical descriptions is of
great importance in this sense. Each statistic has both advantages and
disadvantages. The power spectrum, for example, can fully characterize
the random Gaussian fields, while it does not contain any information
on the non-Gaussianity, which contains significant information in the
gravitational instability theory. Among various statistical
quantities, there is a promising class of statistics which utilizes
smoothed cosmic fields. The smoothed field has less noisy property
than the actual (unsmoothed, or raw) cosmic fields, such as galaxy
distributions, temperature fluctuations of cosmic microwave background
(CMB), shear fields of the gravitational lensing, and so forth.
Perhaps the simplest example of such statistics is the variance
$\langle{\delta_R}^2 \rangle$ of smoothed density contrast $\delta_R$,
which is a function of smoothing length $R$. Similarly, higher-order
cumulants $\langle{\delta_R}^N \rangle_{\rm c}$ ($N=3,4,\ldots$) are
also simple statistics. The density probability distribution function
(PDF) $P(\delta_R)$, which in principle can be constructed from the
hierarchy of cumulants \citep[e.g.,][]{bal89}, is an another example
of popular statistics of smoothed cosmic fields.

Rather recently, more complex statistics of smoothed cosmic fields
have become popular in cosmology, such as the genus statistic
\citep{got86}, density peak statistics \citep{bar86}, area, length,
level-crossing statistics \citep{ryd88a}, Minkowski functionals
\citep{sch97}, etc. These statistics provide assuring
characterizations of the clustering pattern that can not be perceived
only by the hierarchy of cumulants or by the PDF. The genus statistic
is a powerful measure of the morphology in the 3-dimensional
\citep{got89,moo92,par92,bea92,rho94,vog94,pro97,sah97,can98,spr98,col00},
and 2-dimensional \citep{mel89,got92,col91,pli92,dav93,col93,col97}
clustering of galaxies and clusters, and also of the pattern of
temperature fluctuations of CMB radiation
\citep{bon87,tor94,smo94,tor95,par98}. The peak statistics of the
3-dimensional density field are frequently used in connection with the
statistics of the collapsing object
\citep{kai84,man93,cro94,wat94,cen98,gab00}, while the peak statistics
of the CMB fluctuations
\citep{saz85,bon87,col87,vit87,kog95,cay95,fab96,hea99} and of the
weak lensing fields \citep{van00,jai00} are suitable for constraining
cosmological models. The area, length, and level-crossing statistics
directly quantify the amount of contour surfaces
\citep{ryd88b,ryd89,tor94}. The Minkowski functionals
\citep{mec94,sah98,sat98,ker98,bha00}, which are closely related to
the above statistics, are also applied to smoothed cosmic fields
\citep{win97,nas98,sch98,sch99,sch00}. These statistics of smoothed
density fields are considered as powerful descriptors of the
statistical information of the universe. It is therefore essential to
establish theoretical predictions of the behavior of such statistics
so that we can effectively and ideally analyze the data of our
universe.

Recent developments of the perturbation theory \citep[for review,
see][]{ber02} in calculating the variance, the cumulants, and the PDF
are remarkable. The perturbation theory becomes more and more useful
because the recent developments of observations enable us to have
widely covered sample volume of the universe, which can minimize
undesirable strongly nonlinear effects which we do not understand
well. The direct comparison of the theoretical predictions of the
perturbation theory with the actual data is a promising field of
research in this sense. In the case of top-hat smoothing function,
\citet{jus93} and \citet{ber94a} used the tree-level (i.e.,
lowest-order) perturbation theory to obtain the third- and forth-order
cumulants, i.e., the skewness and the kurtosis. The same quantities
with Gaussian smoothing are calculated by \citet{gor86}, \citet{mat94}
and \citet{lok95}. Quite cleverly, \citet{ber94b} took advantage of
special properties of the top-hat smoothing function, and succeeded to
obtain full hierarchy of higher order moments in the tree-level
perturbation theory. He also obtained the PDF from this hierarchy of
cumulants, which remarkably describes the nonlinear behavior of the
gravitational instability in numerical simulations. Beyond the
tree-level calculation, the perturbation theory with loop-corrections
has been also developed
\citep{jus81,vis83,jus84,col90,sut91,mak92,jai94,bau94,sco96a,sco96b}.
Because of the simplicity of the statistics, calculating the variance,
the cumulants and the PDF were primarily the playground of
perturbation-theorists. The evaluation of other statistics by the
perturbation theory is less trivial. A quite useful technique is the
Edgeworth expansion, which was first applied to cosmology in
calculating the PDF from seeds perturbations by \citet{sch91}, and
from non-linear perturbation theory by \citet{jus95},
and \citet{ber95b}. The analytic expression of the genus statistics in
the perturbation theory is derived by \citet{mat94}, whose technique
corresponds to multivariate version of the Edgeworth expansion. In
some literatures the Edgeworth expansion is used in connecting the
statistics and dynamics of the universe \citep{cho97,lok98,tar99}. The
purpose of this paper is to give a comprehensive description of the
formalism by which the perturbative evaluation is possible for wide
range of non-trivial statistics of smoothed cosmic fields in general.

Since the number of spatial dimensions of our universe is three, the
statistics of large-scale cosmic fields are defined in either one-,
two-, or three-dimensional space. For example, the space of the
density field in a redshift map of galaxies or quasars is
three-dimensional. A projected galaxy map on the sky, a shear field of
gravitational lensing, and temperature fluctuations of CMB on the sky
are fields in two-dimensional space. The absorption lines of quasar
spectra and pencil-beam surveys of galaxies define fields in
one-dimensional space. Thus it is useful to develop our statistical
method in $d$-dimensions for generality. As illustrative examples of
applications of our method, we calculate level-crossing statistic (or
equivalently length and area statistics), genus statistics,
density-extrema statistics and Minkowski functionals. As cosmic
fields, we consider density and velocity fields in three dimensions,
the projected density field of galaxies in two dimensions. The basic
formalism and results of the second order theory are presented in this
paper. We will give results of the third order theory in a future
paper, which are technically more involved.

The primary purpose of the present paper is to fully describe the
basic formalism for the statistics of smoothed cosmic fields in
perturbation theory. Applications of the second order theory to
popular statistics and cosmic fields are systematically presented.
Thus, this paper is in some way a mixture of the new results and
comprehensive review of the old results. The major new results in this
paper are:
\begin{itemize}
\item Explicit formulas of the lowest non-Gaussian correction in
various smoothed cosmic fields in $d$-dimensions, including all the
Minkowski functionals. They are presented as functions of both density
threshold and rescaled threshold by volume-fraction.
\item Relations among several statistics of the smoothed field are
clarified.
\item Derivatives of skewness parameter of the velocity field and
projected 2D field are derived in second order perturbation theory of
gravitational instability theory. Especially, skewness parameters with
Gaussian smoothing are detailed. These quantities are particularly
important to perturbatively evaluate the statistics of smoothed field.
\item Genus curve against the scaled threshold in perturbation theory
is discussed and a theory-motivated new asymmetry parameter in genus
curve is introduced.
\end{itemize}

This paper is structured in the following way. \S\ref{sec2} and
\S\ref{sec3} are carrying out a mathematical exercise of expressing
high-order terms in terms of skewness parameters; the physics of
gravity only enters once one actually calculates the skewness
parameters in \S\ref{sec4}. Thus \S\ref{sec2} and \S\ref{sec3} are a
derivation of an extension of the Edgeworth expansion, and the basic
results are equations (\ref{eq1-13}) and (\ref{eq1-17}). In
\S\ref{sec3}, popular statistics of smoothed fields are examined. The
second-order expressions in terms of the skewness parameters are given
for PDF, level-crossing statistics, 2-dimensional and 3-dimensional
genus statistics, 2-dimensional weighted extrema, and Minkowski
functionals in \S\ref{sec3-1}--\S\ref{sec3-6}. These quantities are
re-expressed as functions of the volume-fraction threshold in
\S\ref{sec3-7}. The skewness parameters for several cosmic fields are
calculated in \S\ref{sec4}, applying the second order perturbation
theory. Detailed calculations of the simple hierarchical model, the
3-dimensional density field, the velocity field, the 2-dimensional
projected density field, and the weak lensing field are given in
\S\ref{sec4-1}--\S\ref{sec4-5}. The effect of biasing on the skewness
parameters are discussed in \S\ref{sec4-6}. We discuss implications of
our second-order results in \S\ref{sec5}. We introduce a
theory-motivated asymmetry factor to characterize the shift of the
genus curve in this section. The conclusions are given in
\S\ref{sec6}. Useful Gaussian integrals including Hermite polynomials
are given in Appendix~\ref{app1}. The bispectrum in a two-dimensional
projected field is reviewed in Appendix~\ref{app2}. The symbols in
this paper are summarized in Appendix~\ref{app3}.

\setcounter{equation}{0}
\section{PERTURBATIVE EXPANSION OF STATISTICS}
\label{sec2}

\subsection{Smoothed fields}
\label{sec2-1}

We consider a cosmic random field $f(\bfx)$ which represents any field
constructed from observable quantities of the universe, such as
three-dimensional density field, velocity field, or two-dimensional
projected density field, shear or convergence field of weak lensing,
temperature fluctuations of CMB, etc. The coordinates $\bfx$ can be
either three-, two-, or one-dimensions.

We assume the field $f$ is already smoothed by a smoothing function
$W_R$ with smoothing length $R$ which cuts the high frequency
fluctuations which suffer strongly nonlinear effects:
\begin{eqnarray}
   f(\bfx) = \int d^d x' W_R(|\bfx - \bfx'|) f_{\rm raw}(\bfx'),
\label{eq1-1}
\end{eqnarray}
where $d=1,2,3$ is the dimension of the space $\bfx$, and $f_{\rm
raw}$ is a raw, unsmoothed field. Two of the most popular 3D smoothing
functions are the tophat smoothing function
\begin{eqnarray}
   W_R(x) =  \frac{3}{4\pi R^3} \Theta(R-x),
\label{eq1-1-1}
\end{eqnarray}
and the Gaussian smoothing function
\begin{eqnarray}
   W_R(x) =  \frac{1}{(2\pi)^{3/2} R^3}
   \exp\left(-\frac{x^2}{2R^2}\right).
\label{eq1-1-2}
\end{eqnarray}
We also assume that the mean value of the field $f$ is zero, and that
the existence the variance ${\sigma_0}^2$:
\begin{eqnarray}
   \langle f \rangle = 0,
   \quad
   \langle f^2 \rangle = {\sigma_0}^2.
\label{eq1-2}
\end{eqnarray}
It is convenient to introduce a normalized field $\alpha$ which has a
unit variance as follows:
\begin{eqnarray}
   \alpha \equiv \frac{f}{\sigma_0},
   \quad
   \langle \alpha^2 \rangle = 1.
\label{eq1-3}
\end{eqnarray}

\subsection{Expressing the Non-Gaussian Statistics by Gaussian
   Integration}
\label{sec2-2}

The statistics of a smoothed cosmic field we are interested in are the
functions of the field $\alpha$ and its spatial derivatives as we will
see in the following sections. We denote the series of spatial
derivatives by a set of variables $(A_\mu)$ which is defined by, for
example, in 3D case,
\begin{eqnarray}
   (A_\mu) =
   \left(
      \alpha,
      \partial_1 \alpha,\partial_2 \alpha,\partial_3 \alpha,
      {\partial_1}^2 \alpha,{\partial_2}^2 \alpha,
      {\partial_3}^2 \alpha, 
      \partial_1 \partial_2 \alpha,
      \partial_1 \partial_3 \alpha,
      \partial_2 \partial_3 \alpha,
      \ldots
   \right),
\label{eq1-4}
\end{eqnarray}
where $\partial_i = \partial/\partial x_i$. For convenience, the index
is denoted as $\mu = 0$, 1, 2, 3, (11), (22), (33), (12), (13), (23),
$\ldots$ in such cases. In this example of equation (\ref{eq1-4}),
only the value of the field and of the derivatives on a single point
is considered, but in general, more than two points can be considered.
The set $A_\mu$ forms multivariate random fields, which is denoted as
an $N$-dimensional vector $\bfA$ in the following. The dimension $N$
is the total number of derivatives which appear in the definition of
the statistics we are interested in. The statistical information is
described by the multivariate PDF, $P(\bfA)$. The Fourier transform of
the PDF is the partition function:
\begin{eqnarray}
   Z(\bfJ) = 
   \int_{-\infty}^\infty d^N\!A\, P(\bfA)
   \exp ( i \bfJ \cdot \bfA ).
\label{eq1-5}
\end{eqnarray}
At this point, the cumulant expansion theorem \citep[e.g.,][]{ma85} is
very useful. This theorem states that $\ln Z$ is the generating
function of the cumulants, $M^{(n)}_{\mu_1 \cdots \mu_n} \equiv
\langle A_{\mu_1} \cdots A_{\mu_n} \rangle_{\rm c}$:
\begin{eqnarray}
   \ln Z(\bfJ) =
   \sum_{n=1}^\infty \frac{i^n}{n!}
   \sum_{\mu_1 = 1}^N \cdots \sum_{\mu_n = 1}^N
   M^{(n)}_{\mu_1 \cdots \mu_n}
   J_{\mu_1} \cdots J_{\mu_n}.
\label{eq1-6}
\end{eqnarray}
It follows from $\langle f \rangle = 0$ that $\langle A_\mu \rangle =
0$, and first several cumulants are given by
\begin{eqnarray}
&&
   M^{(1)}_{\mu} = 0,
\label{eq1-7a}\\
&&
   M^{(2)}_{\mu_1 \mu_2} =
   \left\langle A_{\mu_1} A_{\mu_2} \right\rangle,
\label{eq1-7b}\\
&&
   M^{(3)}_{\mu_1 \mu_2 \mu_3} =
   \left\langle A_{\mu_1} A_{\mu_2} A_{\mu_3} \right\rangle,
\label{eq1-7c}\\
&&
   M^{(4)}_{\mu_1 \mu_2 \mu_3 \mu_4} =
   \left\langle
      A_{\mu_1} A_{\mu_2} A_{\mu_3} A_{\mu_4}
   \right\rangle -
   \left\langle A_{\mu_1} A_{\mu_2} \right\rangle
   \left\langle A_{\mu_3} A_{\mu_4} \right\rangle
-
   \left\langle A_{\mu_1} A_{\mu_3} \right\rangle
   \left\langle A_{\mu_2} A_{\mu_4} \right\rangle -
   \left\langle A_{\mu_1} A_{\mu_4} \right\rangle
   \left\langle A_{\mu_2} A_{\mu_3} \right\rangle,
\label{eq1-7d}
\end{eqnarray}
and so forth. {}From equations (\ref{eq1-6}) and (\ref{eq1-7a}), the
partition function is given by
\begin{eqnarray}
   Z(\bfJ) = 
   \exp\left(
      -\frac12 \bfJ^{\rm T} \bfM \bfJ \right)
   \exp\left(
      \sum_{n=3}^\infty \frac{i^n}{n!}
      \sum_{\mu_1,\cdots \mu_n}
      M^{(n)}_{\mu_1 \cdots \mu_n} J_{\mu_1} \cdots J_{\mu_n} \right),
\label{eq1-8}
\end{eqnarray}
where $\bfM$ is an $N\times N$ matrix whose components are given by
$M^{(2)}_{\mu\nu}$. On the other hand, the equation (\ref{eq1-5}) is
inverted as
\begin{eqnarray}
   P(\bfA) = \frac{1}{(2\pi)^N}
   \int_{-\infty}^\infty d^N\!J\, Z(\bfJ)
   \exp \left( -i \bfJ\cdot\bfA \right).
\label{eq1-9}
\end{eqnarray}
Substituting $J_\mu \rightarrow i\partial/\partial A_\mu$ in the last
term of equation (\ref{eq1-8}), the distribution function of equation
(\ref{eq1-9}) can be transformed in a form
\begin{eqnarray}
   P(\bfA) = 
   \exp\left(
      \sum_{n=3}^\infty \frac{(-)^n}{n!}
      \sum_{\mu_1,\cdots \mu_n}
      M^{(n)}_{\mu_1 \cdots \mu_n} 
      \frac{\partial^n}
         {\partial A_{\mu_1} \cdots \partial A_{\mu_n}}
   \right)
   P_{\rm G}(\bfA),
\label{eq1-10}
\end{eqnarray}
where
\begin{eqnarray}
   P_{\rm G}(\bfA) &=&
   \frac{1}{(2\pi)^N}
   \int_{-\infty}^\infty d^N\!J\, 
   \exp \left(
   - i \bfJ\cdot\bfA 
   - \frac12 \bfJ^{\rm T} \bfM \bfJ
   \right)
\label{eq1-11a}\\
   &=&
   \frac{1}{(2\pi)^{N/2} \sqrt{\det\bfM}}
   \exp\left(-\frac12 \bfA^{\rm T} \bfM^{-1} \bfA \right),
\label{eq1-11b}
\end{eqnarray}
is the multivariate Gaussian distribution function characterized by
the correlation matrix $\bfM$.

Any statistical quantity of a smoothed cosmic field is expressed by an
average $\langle F \rangle$ of a certain function $F(\bfA)$ as we will
see in the following sections. Thus, from equation (\ref{eq1-10}), we
obtain
\begin{eqnarray}
   \langle F \rangle &=&
   \int_{\infty}^\infty d^N\!A\,P(\bfA) F(\bfA)
\label{eq1-12a}\\
   &=&
   \left\langle
   \exp\left(
      \sum_{n=3}^\infty \frac{1}{n!}
      \sum_{\mu_1,\cdots \mu_n}
      M^{(n)}_{\mu_1 \cdots \mu_n} 
      \frac{\partial^n}
         {\partial A_{\mu_1} \cdots \partial A_{\mu_n}}
   \right)
   F(\bfA)
   \right\rangle_{\rm G},
\label{eq1-12b}
\end{eqnarray}
where
\begin{eqnarray}
   \langle \cdots \rangle_{\rm G} \equiv
   \int_{\infty}^\infty d^N\!A\,P_{\rm G}(\bfA) F(\bfA)
\label{eq1-12-1}
\end{eqnarray}
denotes the averaging by the Gaussian distribution function of
equation (\ref{eq1-11b}).

This form, equation (\ref{eq1-12b}), is useful when the deviation from
the Gaussian distribution is not large. In principle, this equation
reduces the general statistical averaging procedure to Gaussian
integrations. However, it contains the infinite series, thus we have
to truncate this expression by some criteria. In most cases of
interest, the weakly nonlinear evolution of cosmic fields satisfy
$M^{(n)} \sim {\cal O}({\sigma_0}^{n-2})$, as we will see in
\S~\ref{sec4}. When this relation holds, we can expand the
distribution function to arbitrary order in $\sigma_0$. In the
following, we assume this relation, and introduce the normalized
cumulants:
\begin{eqnarray}
   \widehat{M}^{(n)}_{\mu_1\cdots\mu_n} = 
   \frac{M^{(n)}_{\mu_1\cdots\mu_n}}{{\sigma_0}^{n-2}},
\label{eq1-12}
\end{eqnarray}
which are assumed to be of order one in terms of $\sigma_0$. In this
case, the equation (\ref{eq1-12b}) is expanded as, up to ${\cal
O}({\sigma_0}^2)$,
\begin{eqnarray}
&&
   \langle F \rangle =
   \langle F \rangle_{\rm G} +
   \frac{1}{3!}\sum \hMt_{\mu_1 \mu_2 \mu_3}
   \left\langle F_{,\mu_1 \mu_2 \mu_3} \right\rangle_{\rm G} \sigma_0
\nonumber\\
&& \qquad\qquad
   \, +
   \left[
      \frac{1}{4!} \sum \widehat{M}^{(4)}_{\mu_1 \mu_2 \mu_3 \mu_4}
      \left\langle
      F_{,\mu_1\mu_2\mu_3\mu_4} \right\rangle_{\rm G} +
      \frac{1}{2\cdot (3!)^2} \sum \widehat{M}^{(3)}_{\mu_1\mu_2\mu_3}
      \widehat{M}^{(3)}_{\nu_1\nu_2\nu_3}
      \left\langle
      F_{,\mu_1 \mu_2 \mu_3 \nu_1 \nu_2 \nu_3}
      \right\rangle_{\rm G}
   \right] {\sigma_0}^2
   + {\cal O}({\sigma_0}^3),
\label{eq1-13}
\end{eqnarray}
where we introduce the notation, $F_{,\mu_1\mu_2\mu_3} \equiv
\partial^3 F/\partial A_{\mu_1} \partial A_{\mu_2}\partial A_{\mu_3}$
etc. The calculation of the factors $\langle F_{,\mu\cdots}
\rangle_{\rm G}$ is performed for individual statistic which gives the
explicit form of the function $F$.

\subsection{Two-point Correlations}
\label{sec2-3}

In the expansion of equation (\ref{eq1-13}), we need to calculate the
Gaussian average of derivatives of the function $F$:
\begin{eqnarray}
&&
   \langle F_{,\mu_1\mu_2\cdots} \rangle_{\rm G} =
   \int d^N A P_{\rm G}(\bfA) F_{,\mu_1\mu_2\cdots}(\bfA)
   =
   \frac{1}{(2\pi)^{N/2} \sqrt{\det\bfM}}
   \int d^N A
   \exp\left(-\frac12 \bfA^{\rm T} \bfM^{-1} \bfA \right)
    F_{,\mu_1\mu_2\cdots}(\bfA).
\label{eq1-13-1}   
\end{eqnarray}
In most cases, the evaluation is
analytically feasible because only Gaussian integration is needed. For
the evaluation of such integration, we need the correlation matrix
$\bfM$. Throughout this paper, we consider statistically homogeneous,
and isotropic fields, in which case, the correlation matrix $\bfM$ is
simplified because of the symmetry. In fact, the correlation matrix
takes the following forms:
\begin{eqnarray}
&&
   \langle \alpha \alpha \rangle = 1,
\label{eq1-13-2a}\\
&&
   \langle \alpha \alpha_{,i} \rangle = 0,
\label{eq1-13-2b}\\
&&
   \langle \alpha \alpha_{,ij} \rangle =
   - \frac{1}{d} \frac{{\sigma_1}^2}{{\sigma_0}^2} \delta_{ij},
\label{eq1-13-2c}\\
&&
   \langle \alpha_{,i} \alpha_{,j} \rangle =
   \frac{1}{d} \frac{{\sigma_1}^2}{{\sigma_0}^2} \delta_{ij},
\label{eq1-13-2d}\\
&&
   \langle \alpha_{,i} \alpha_{,jk} \rangle = 0,
\label{eq1-13-2e}\\
&&
   \langle \alpha_{,ij} \alpha_{,kl} \rangle = 
   \frac{1}{d(d+2)} \frac{{\sigma_2}^2}{{\sigma_0}^2}
   \left(
      \delta_{ij}\delta_{kl} +
      \delta_{ik}\delta_{jl} +
      \delta_{il}\delta_{jk}
   \right),
\label{eq1-13-2f}
\end{eqnarray}
where $\alpha_{,ij} = \partial\alpha/\partial x_i\partial x_j$ etc.
and we define the following quantities:
\begin{eqnarray}
&&
   {\sigma_1}^2 = 
   - \langle f \nabla^2 f \rangle =
   - \langle \alpha \nabla^2 \alpha \rangle {\sigma_0}^2.
\label{eq1-13-3a}\\
&&
   {\sigma_2}^2 = 
   \langle \nabla^2 f \nabla^2 f \rangle =
   \langle \nabla^2 \alpha \nabla^2 \alpha \rangle {\sigma_0}^2.
\label{eq1-13-3b}
\end{eqnarray}
In most cases of interest, the derivatives of order higher than two do
not appear in the definition of the statistics, so we do not give
explicit form of those correlations.

In the following, we use the notation $\eta_i = \alpha_{,i}$ and
$\zeta_{ij} = \alpha_{,ij}$ following \citet{bar86}. As is often the
case, when the second order derivatives $\zeta_{ij}$ appears in $F$ as
simple polynomials, the following transform is particularly useful:
\begin{eqnarray}
   \widetilde{\zeta}_{ij} = 
   \zeta_{ij} +
   \frac{1}{d}\frac{{\sigma_1}^2}{{\sigma_0}^2}
   \delta_{ij}
   \alpha.
\label{eq1-13-3-1}
\end{eqnarray}
This transform erase the correlation between $\alpha$ and
$\widetilde{\zeta}_{ij}$, and the non-zero correlations are only
\begin{eqnarray}
&&
   \langle\alpha^2\rangle = 1,
\label{eq1-13-3-2a}\\
&&
   \langle{\eta_1}^2\rangle =
   \langle{\eta_2}^2\rangle =
   \langle{\eta_3}^2\rangle =
   \frac{1}{d} \frac{{\sigma_1}^2}{{\sigma_0}^2},
\label{eq1-13-3-2b}\\
&&
   \langle{\widetilde{\zeta}_{11}}^2\rangle =
   \langle{\widetilde{\zeta}_{22}}^2\rangle =
   \langle{\widetilde{\zeta}_{33}}^2\rangle =
   \frac{3}{d(d+2)} \frac{{\sigma_2}^2}{{\sigma_0}^2}
   \left(
      1 - \frac{d+2}{3d} \gamma^2
   \right),
\label{eq1-13-3-2c}\\
&&
   \langle\widetilde{\zeta}_{11}\widetilde{\zeta}_{22}\rangle =
   \langle\widetilde{\zeta}_{11}\widetilde{\zeta}_{33}\rangle =
   \langle\widetilde{\zeta}_{22}\widetilde{\zeta}_{33}\rangle =
   \frac{1}{d(d+2)} \frac{{\sigma_2}^2}{{\sigma_0}^2}
   \left(
      1 - \frac{d+2}{d} \gamma^2
   \right),
\label{eq1-13-3-2d}\\
&&
   \langle{\widetilde{\zeta}_{12}}^2\rangle =
   \langle{\widetilde{\zeta}_{13}}^2\rangle =
   \langle{\widetilde{\zeta}_{23}}^2\rangle =
   \frac{1}{d(d+2)} \frac{{\sigma_2}^2}{{\sigma_0}^2},
\label{eq1-13-3-2e}
\end{eqnarray}
where
\begin{eqnarray}
   \gamma = \frac{{\sigma_1}^2}{\sigma_0\sigma_2}.
\label{eq1-13-5}
\end{eqnarray}
The Gaussian integration is straightforward if the function $F$ is
expressed by polynomials of $\widetilde{\zeta}_{ij}$.

If the function $F$ is more complicated in which $\zeta_{ij}$ are not
simply given by polynomials, it is useful to completely diagonalize
the correlation matrix $\bfM$ of equations
(\ref{eq1-13-2a})--(\ref{eq1-13-2a}). We introduce the following
transform, which is quite similar one in \citet{bar86}:
\begin{eqnarray}
&&
   x =
   - \frac{\sigma_0}{\sigma_2}
   \left(
      \sum_i \zeta_{ii}
      + \frac{{\sigma_1}^2}{{\sigma_0}^2}\alpha
   \right),
\label{eq1-13-4a}\\
&&
   y =
   - \frac{\sigma_0}{\sigma_2}
   \frac{\zeta_{11} - \zeta_{22},}{2}
\label{eq1-13-4b}\\
&&
   z =
   - \frac{\sigma_0}{\sigma_2}
   \frac{\zeta_{11} + \zeta_{22} - 2\zeta_{33}}{2}.
\label{eq1-13-4c}
\end{eqnarray}
If $d=1$, we ignore the variables $y,z$, and similarly, if $d=2$, we
ignore the variable $z$ in the above equations and in the following.
The above transform is similar to \citet{bar86} but notice it is not
identical. This transform completely diagonalize the correlation
matrix:
\begin{eqnarray}
&&
   \langle \alpha^2 \rangle = 1,
\label{eq1-13-6a}\\
&&
   \langle {\eta_1}^2 \rangle =
   \langle {\eta_2}^2 \rangle =
   \langle {\eta_3}^2 \rangle =
   \frac{1}{d} \frac{{\sigma_1}^2}{{\sigma_0}^2},
\label{eq1-13-6b}\\
&&
   \langle x^2 \rangle = 1 - \gamma^2,
\label{eq1-13-6c}\\
&&
   \langle y^2 \rangle = \frac{1}{d(d+2)},
\label{eq1-13-6d}\\
&&
   \langle z^2 \rangle = \frac{3}{d(d+2)},
\label{eq1-13-6e}\\
&&
   \langle {\zeta_{12}}^2 \rangle =
   \langle {\zeta_{13}}^2 \rangle =
   \langle {\zeta_{23}}^2 \rangle =
   \frac{1}{d(d+2)} \frac{{\sigma_2}^2}{{\sigma_0}^2},
\label{eq1-13-6f}
\end{eqnarray}
and all non-diagonal correlations are zero. For later convenience, we
write the inverse relation of the transform of $x$, $y$, and $z$:
\begin{eqnarray}
&&
   \zeta_{11} =
   - \frac{\sigma_2}{4\sigma_0}
   (x + 4y + 2z + \gamma\alpha),
\label{eq1-13-7a}\\
&&
   \zeta_{22} =
   - \frac{\sigma_2}{4\sigma_0}
   (x - 4y + 2z + \gamma\alpha),
\label{eq1-13-7b}\\
&&
   \zeta_{33} =
   - \frac{\sigma_2}{2\sigma_0}
   (x - 2z + \gamma\alpha),
\label{eq1-13-7c}
\end{eqnarray}
If $\zeta_{33}$ does not appear in the function $F$, the variable
$z$ should be omitted in the above equations. If both $\zeta_{22}$
and $\zeta_{33}$ do not appear in the function $F$, the variables
$y$ and $z$ should be omitted.

After expressing the function $F$ in terms of the diagonalized
variables $\alpha$, $\eta_i$, $x$, $y$, $z$ and $\zeta_{ij}\ (i<j)$,
the calculation of the Gaussian integration of equation
(\ref{eq1-13-1}) is performed.

\subsection{Three-point Correlations}
\label{sec2-4}

In this paper, only first two terms in equation (\ref{eq1-13}) are
considered. Thus we evaluate $M^{(3)}$ here. When the spatial symmetry
is taken into account, the quantity $M^{(3)}$ reduces to the following 
expressions:
\begin{eqnarray}
&&
   \langle \alpha^2 \alpha_{,i} \rangle = 0,
\label{eq1-14a}\\
&&
   \langle \alpha^2 \alpha_{,ij} \rangle = 
   \frac{\langle \alpha^2 \nabla^2 \alpha \rangle}{d} 
   \delta_{ij},
\label{eq1-14b}\\
&&
   \langle \alpha \alpha_{,i} \alpha_{,j} \rangle =
   - \frac{\langle \alpha^2 \nabla^2 \alpha \rangle}{2d} \delta_{ij},
\label{eq1-14c}\\
&&
   \langle \alpha \alpha_{,i} \alpha_{,jk} \rangle = 0,
\label{eq1-14d}\\
&&
   \langle \alpha \alpha_{,ij} \alpha_{,kl} \rangle = 
   \frac{\langle\alpha\nabla^2\alpha\nabla^2\alpha\rangle}
      {d(d+2)}
   \left(
      \delta_{ij}\delta_{kl} + \delta_{ik}\delta_{jl} + 
      \delta_{il}\delta_{jk}
   \right)
   -
   \frac{3
      \langle(\nabla\alpha\cdot\nabla\alpha)\nabla^2\alpha\rangle}
      {d(d-1)(d+2)}
   \left[
      \delta_{ij}\delta_{kl} -
      \frac{d}{2}
      \left(
         \delta_{ik}\delta_{jl} + \delta_{il}\delta_{jk}
      \right)
   \right],
\label{eq1-14e}\\
&&
   \langle\alpha_{,i}\alpha_{,j}\alpha_{,k}\rangle = 0,
\label{eq1-14f}\\
&&
   \langle\alpha_{,i}\alpha_{,j}\alpha_{,kl}\rangle = 
   \frac{\langle(\nabla\alpha\cdot\nabla\alpha)\nabla^2\alpha\rangle}
      {d(d-1)}
   \left[
      \delta_{ij}\delta_{kl} -
      \frac12
      \left(\delta_{ik}\delta_{jl} + \delta_{il}\delta_{jk}\right)
   \right],
\label{eq1-14g}\\
&&
   \langle\alpha_{,i}\alpha_{,jk}\alpha_{,lm}\rangle = 0,
\label{eq1-14h}
\end{eqnarray}
etc. To prove the above equations, the following identities for
isotropic fields are useful:
\begin{eqnarray}
&&
   \langle\alpha_{,i}\alpha_{,j}\alpha_{,ij}\rangle = 
   - \frac12 
   \langle\nabla\alpha\cdot\nabla\alpha)\nabla^2\alpha\rangle,
\label{eq1-14-1a}\\
&&
   \langle\alpha\alpha_{,ij}\alpha_{,ij}\rangle = 
   \frac32 
   \langle\nabla\alpha\cdot\nabla\alpha)\nabla^2\alpha\rangle +
   \langle\alpha(\nabla^2\alpha)(\nabla^2\alpha)\rangle.
\label{eq1-14-1b}
\end{eqnarray}
Although the above equations are valid for $d \ne 0, 1$, the case
$d=1$ is obtained by just ignoring the terms with $(d-1)^{-1}$.
Generally, more complicated quantities can be appeared for $M^{(3)}$,
but the above relations are sufficient for our applications in this
paper.

According to the spatial symmetry of the above equations, the
third-order correlations $\hMt_{\mu\nu\lambda}$ are explicitly given.
At this point, it is useful to define the following quantities:
\begin{eqnarray}
&&
   S^{(0)} =
   \frac{\langle f^3 \rangle}{{\sigma_0}^4} =
   \frac{\langle \alpha^3 \rangle}{\sigma_0},
\label{eq1-15a}\\
&&
   S^{(1)} = - \frac34 \frac{\langle f^2 (\nabla^2 f) \rangle}
                {{\sigma_0}^2{\sigma_1}^2} =
   - \frac34 \frac{\langle \alpha^2 (\nabla^2 \alpha) \rangle \sigma_0}
        {{\sigma_1}^2},
\label{eq1-15b}\\
&&
   S^{(2)} = - \frac{3d}{2(d-1)}
      \frac{\langle(\nabla f\cdot\nabla f)(\nabla^2 f) \rangle}
         {{\sigma_1}^4} =
   - \frac{3d}{2(d-1)}
     \frac{\langle(\nabla\alpha\cdot\nabla\alpha)
         (\nabla^2\alpha)\rangle {\sigma_0}^3}
        {{\sigma_1}^4},
\label{eq1-15c}\\
&&
   S^{(2)}_2 = \frac{\langle f(\nabla^2 f)(\nabla^2 f)\rangle}
                {{\sigma_1}^4} =
   \frac{\langle\alpha(\nabla^2\alpha)(\nabla^2\alpha)\rangle
         {\sigma_0}^3}
        {{\sigma_1}^4}.
\label{eq1-15d}
\end{eqnarray}
We call quantities $S^{(a)}$ skewness parameters. The first one
$S^{(0)}$ is usually called as skewness, and the others are its
derivatives. They are to be calculated from usual perturbation
theories in \S~\ref{sec4}. Using these quantities, the third-order
correlations are given by
\begin{eqnarray}
&&
   \hMt_{000} = S^{(0)},
\label{eq1-16a}\\
&&
   \hMt_{00i} = 0,
\label{eq1-16b}\\
&&
   \hMt_{00(ii)} =
   - \frac{4}{3d}\frac{{\sigma_1}^2}{{\sigma_0}^2} S^{(1)},
\label{eq1-16c}\\
&&
   \hMt_{00(ij)} = 0, \qquad (i < j),
\label{eq1-16d}\\
&&
   \hMt_{0ii} =
   \frac{2}{3d} \frac{{\sigma_1}^2}{{\sigma_0}^2}
   S^{(1)},
\label{eq1-16e}\\
&&
   \hMt_{0ij} = 0, \qquad (i \ne j),
\label{eq1-16f}\\
&&
   \hMt_{0i(jk)} = 0,
\label{eq1-16g}\\
&&
   \hMt_{0(ii)(ii)} =
   -\frac{1}{d^2 (d+2)} \frac{{\sigma_1}^4}{{\sigma_0}^4}
   \left[2(d-1) S^{(2)} - 3d S^{(2)}_2 \right),
\label{eq1-16h}\\
&&
   \hMt_{0(ii)(jj)} =
   \frac{1}{d^2 (d+2)}\frac{{\sigma_1}^4}{{\sigma_0}^4}
   \left[ 2 S^{(2)} + d S^{(2)}_2 \right],
   \qquad (i \ne j),
\label{eq1-16i}\\
&&
   \hMt_{0(ii)(jk)} = 0,
   \qquad (j < k),
\label{eq1-16j}\\
&&
   \hMt_{0(ij)(ij)} = 
   - \frac{1}{d(d+2)}\frac{{\sigma_1}^4}{{\sigma_0}^4}
   \left[ S^{(2)} -  S^{(2)}_2 \right]
   \qquad (i < j),
\label{eq1-16k}\\
&&
   \hMt_{0(ij)(kl)} = 0,
   \qquad (i < j, k < l, i \ne k),
\label{eq1-16l}\\
&&
   \hMt_{ijk} = 0,
\label{eq1-16m}\\
&&
   \hMt_{ii(ii)} = 0,
\label{eq1-16n}\\
&&
   \hMt_{ii(jj)} =
   - \frac{2}{3 d^2}\frac{{\sigma_1}^4}{{\sigma_0}^4} S^{(2)},
   \qquad (i \ne j),
\label{eq1-16o}\\
&&
   \hMt_{ii(jk)} = 0,
   \qquad (j < k),
\label{eq1-16p}\\
&&
   \hMt_{ij(kk)} = 0,
   \qquad (i \ne j),
\label{eq1-16q}\\
&&
   \hMt_{ij(ij)} =
   \frac{1}{3 d^2}\frac{{\sigma_1}^4}{{\sigma_0}^4} S^{(2)},
   \qquad (i < j),
\label{eq1-16r}\\
&&
   \hMt_{ij(kl)} = 0,
   \qquad (i < j, k < l, i \ne k),
\label{eq1-16s}\\
&&
   \hMt_{i(jk)(lm)} = 0,
\label{eq1-16t}
\end{eqnarray}
and so forth, where repeated indices are {\em not} summed over in the
above equations, and $\hMt$ is symmetric under permutation of its
indices. Thus, denoting $F_{\mu\nu\lambda} \equiv \langle
F_{,\mu\nu\lambda} \rangle_{\rm G}$ for simplicity,
\begin{eqnarray}
&&
   \sum_{\mu,\nu,\lambda} \hMt_{\mu\nu\lambda} F_{\mu\nu\lambda} =
   \hMt_{000} F_{000} +
   3 \sum_i \hMt_{00(ii)} F_{00(ii)} +
   3 \sum_i \hMt_{0ii} F_{0ii} +
   3 \sum_i \hMt_{0(ii)(ii)} F_{0(ii)(ii)}
\nonumber\\
&&\qquad\qquad +\,
   6 \sum_{i<j} \hMt_{0(ii)(jj)} F_{0(ii)(jj)} +
   3 \sum_{i<j} \hMt_{0(ij)(ij)} F_{0(ij)(ij)} +
   3 \sum_{i\ne j} \hMt_{ii(jj)} F_{ii(jj)} +
   6 \sum_{i<j} \hMt_{ij(ij)} F_{ij(ij)} + 
   \cdots
\nonumber\\
&&\qquad =
   S^{(0)} F_{000} -
   \frac{2 S^{(1)}}{d}
   \frac{{\sigma_1}^2}{{\sigma_0}^2}
   \left(
      2 \sum_i F_{00(ii)} - \sum_i F_{0ii}
   \right)
\nonumber\\
&&\quad\qquad -\,
   \frac{2 S^{(2)}}{d^2 (d+2)}
   \frac{{\sigma_1}^4}{{\sigma_0}^4}
   \left[
      3(d-1) \sum_i F_{0(ii)(ii)} -
      6 \sum_{i<j} F_{0(ii)(jj)}
   +
      \frac{3d}{2} \sum_{i<j} F_{0(ij)(ij)} +
      (d+2) \sum_{i\ne j} F_{ii(jj)} -
      (d+2) \sum_{i<j} F_{ij(ij)}
   \right] 
\nonumber\\
&&\quad\qquad +\,
   \frac{3 S^{(2)}_2}{d(d+2)}
   \frac{{\sigma_1}^4}{{\sigma_0}^4}
   \left(
      3 \sum_{i} F_{0(ii)(ii)} +
      2 \sum_{i<j} F_{0(ii)(jj)} +
      \sum_{i<j} F_{0(ij)(ij)}
   \right) + \cdots.
\label{eq1-17}
\end{eqnarray}
This equation gives the second-order correction term of the
statistical quantity $\langle F \rangle$ through equation
(\ref{eq1-13}). Once the field $f$ is specified, the skewness
parameters $S^{(a)}$ are calculated by dynamical perturbation theory
of the field $f$. The remaining factors in the above equation are the
Gaussian integrations of the derivatives of the function $F$, i.e.,
$\langle F_{,\mu\nu\lambda} \rangle_{\rm G}$. These factors can be
calculated once the function $F$ is given. In the next section, we
calculate the latter factors for individual statistics.

\setcounter{equation}{0}
\section{STATISTICS OF SMOOTHED COSMIC FIELD}
\label{sec3}

In this section we calculate the factor $\langle F_{,\mu_1\mu_2\cdots}
\rangle_{\rm G}$ for each statistic. Some of the results in this
section have been previously presented. The Edgeworth expansion of the
PDF in \S\ref{sec3-1} is a familiar result. The result of 3D genus
statistic in \S\ref{sec3-4} was already given by \citet{mat94}. We
include these old results for completeness. Other subsections present
new results.

\subsection{Probability Distribution Function}
\label{sec3-1}

Perhaps the simplest yet non-trivial statistic is the PDF, $P(f)$. The
perturbative expansion of the PDF is known as the Edgeworth expansion
\citep{sch91,jus95,ber95b}. As the simplest example, we re-derive the
known Edgeworth expansion from the point of view of our general
formalism above \citep[see also][]{mat95a}.

Since the PDF is simply given by $P(f) = \langle\delta(f' -
f)\rangle_{f'}$, where $\delta$ is the Dirac's delta function, the
function $F$ in the previous section for PDF $P(f)$ is given by
\begin{eqnarray}
   F =
   \frac{1}{\sigma_0} \delta(\alpha - f/\sigma_0)
\label{eq2-0-1}
\end{eqnarray}
Since this form of $F$ does not depend on derivatives of $\alpha$,
only $F_{000}$ survives in the equation (\ref{eq1-17}). {}From
equation (\ref{eqa1-4a}) with $n=0$ and $k=3$, $F_{000} = (2\pi)^{1/2}
e^{-\nu^2/2} H_3(\nu)$, where $\nu = f/\sigma_0$. Thus, the PDF is
derived from the equation (\ref{eq1-13}):
\begin{eqnarray}
   P(f) = 
   \frac{e^{-\nu^2/2}}{\sqrt{2\pi}\sigma_0}
   \left[
      1 + \sigma_0 \frac{S^{(0)}}{6} H_3(\nu)
      + {\cal O}({\sigma_0}^2)
   \right],
\label{eq2-0-2}
\end{eqnarray}
which reproduces the well-known result.

There is less advantage of applying our formalism to this simple
statistic which can be treated by standard methods. Our formalism has
advantages when more non-trivial statistics are considered as shown
below.

\subsection{Level Crossing, Length and Area Statistics}
\label{sec3-2}

Next three statistics we consider here are the level-crossing
statistic $N_1$, the length statistic $N_2$ and the area statistic
$N_3$. The level-crossing statistic is defined by the mean number of
intersection of a straight line and threshold contours of the field.
The length statistic is defined by the mean length of intersection of
a 2D surface and the threshold contours of the field. The area
statistic is defined by the mean area of the contour surface in a 3D
space \citep{ryd88a,ryd89,mat96a}. The level-crossing statistic is
defined for 1D, 2D, and 3D cosmic fields, the length statistic is
defined for 2D and 3D cosmic fields, while the area statistic is
defined only for 3D cosmic fields. For statistically isotropic fields,
those three statistics are proportional to each other.

In general, a statistic of the smoothed field $f$ is a function of the
threshold $f_{\rm t}$, or of the normalized threshold $\nu = f_{\rm
t}/\sigma_0$. The explicit expressions of statistics $N_1$, $N_2$, and
$N_3$ are given by \citep{ryd88a}
\begin{eqnarray}
&&
   N_1(\nu) = 
   \left\langle
      \delta(\alpha - \nu)\left|\eta_1\right|
   \right\rangle,
\label{eq2-1a}\\
&&
   N_2(\nu) = 
   \left\langle
      \delta(\alpha - \nu)
      \left[
         (\eta_1)^2 + (\eta_2)^2
      \right]^{1/2}
   \right\rangle,
\label{eq2-1b}\\
&&
   N_3(\nu) = 
   \left\langle
      \delta(\alpha - \nu)
      \left[
         (\eta_1)^2 + (\eta_2)^2 + (\eta_3)^2
      \right]^{1/2}
   \right\rangle.
\label{eq2-1c}
\end{eqnarray}
For isotropic fields, these statistics are actually equivalent
\citep{ryd88a}. In fact, the distribution function of $\eta_i \equiv
\alpha_{,i}$ for fixed $\alpha = \nu$ is the function of only the
magnitude $|\bfeta|$. Thus, using spherical coordinates for $\bfeta$,
one can see
\begin{eqnarray}
&&
   N_1(\nu) =
   N_2(\nu) \int\frac{d\phi}{2\pi} |\cos\phi| =
   N_3(\nu) \int\frac{d\Omega}{4\pi} |\sin\theta \cos\phi|,
\label{eq2-2b}
\end{eqnarray}
i.e.,
\begin{eqnarray}
   N_1(\nu) = \frac{2}{\pi} N_2(\nu) = \frac12 N_3(\nu).
\label{eq2-2}
\end{eqnarray}
Thus we only need to consider $N_1$ which has the simplest expression
and the rests of the statistics are automatically given by equation
(\ref{eq2-2}). However, if the field is anisotropic, such as the
density field in redshift space \citep{mat96a}, the equation
(\ref{eq2-2}) no longer holds, and equations
(\ref{eq2-1a})--(\ref{eq2-1c}) should be used for each statistic. The
equation (\ref{eq1-17}) only holds for statistically isotropic fields,
but the equation (\ref{eq1-13}) is applicable even for anisotropic
fields.

Now we calculate the factor $\langle F_{,\mu_1\mu_2\cdots}\rangle_{\rm
G}$ for the particular statistic $N_1$. The indices $\mu_1, \mu_2 ,
\ldots$ only take $0$ and $1$ for $N_1$ statistic. Let the number of
$0$ be $k$ and the number of $1$ be $l$. Then the factor is given by
\begin{eqnarray}
   \left\langle F_{,\mu_1\mu_2\cdots}\right\rangle_{\rm G} =
   R(k,l) \equiv
   \left\langle
      \left(\frac{\partial}{\partial\alpha}\right)^k
      \left(\frac{\partial}{\partial\eta_1}\right)^l
      \delta(\alpha - \nu) |\eta_1|
   \right\rangle_{\rm G}.
\label{eq2-3}
\end{eqnarray}
Since the variables $\alpha$ and $\eta_i$ are uncorrelated in the
Gaussian averaging, we can use equations (\ref{eqa1-4a}) and
(\ref{eqa1-5a}) in appendix~\ref{app1}. Thus the above Gaussian
integration results in
\begin{eqnarray}
   R(k,l) =
   \frac{h_{l-2}}{\pi}
   \left(
      \frac{\sigma_1}{\sqrt{d}\sigma_0}
   \right)^{1-l}
   e^{-\nu^2/2} H_k(\nu),
\label{eq2-4}
\end{eqnarray}
where $h_l$ is given by equation (\ref{eqa1-3}). Now, calculation of
equation (\ref{eq1-17}) is straightforward:
\begin{eqnarray}
&&
   \sum_{\mu,\nu,\lambda} \hMt_{\mu\nu\lambda} F_{\mu\nu\lambda} =
   S^{(0)} R(3,0) +
   2 S^{(1)} \left(\frac{\sigma_1}{\sqrt{d}\sigma_0}\right)^2
   R(1,2)
   =
   \frac{1}{\pi} \frac{\sigma_1}{\sqrt{d}\sigma_0}
   e^{-\nu^2/2}
   \left[
      S^{(0)} H_3(\nu) + 2 S^{(1)} H_1(\nu)
   \right].
\label{eq2-5}
\end{eqnarray}
On the other hand, the Gaussian contribution is simply given by
\begin{eqnarray}
   \langle F \rangle_{\rm G} =
   R(0,0) =
   \frac{1}{\pi}\frac{\sigma_1}{\sqrt{d}\sigma_0} e^{-\nu^2/2}.
\label{eq2-6}
\end{eqnarray}
Thus, the perturbative expansion of equation (\ref{eq1-13}) up to
second order is finally given by
\begin{eqnarray}
   N_1(\nu) =
   \frac{1}{\pi} \frac{\sigma_1}{\sqrt{d}\sigma_0} e^{-\nu^2/2}
   \left\{
      1 +
      \left[
         \frac{S^{(0)}}{6} H_3(\nu) +
         \frac{S^{(1)}}{3} H_1(\nu)
      \right] \sigma_0 +
      {\cal O}({\sigma_0}^2)
   \right\}.
\label{eq2-7}
\end{eqnarray}
The second order formulas for area and length statistics are given by
equation (\ref{eq2-2}) with the above equation. To evaluate the above
formula, the factor $S^{(0)}$ and $S^{(1)}$ should be known. Those
factors are evaluated by usual perturbation theory in the following
section.

\subsection{2D Genus Statistic}
\label{sec3-3}

The next statistics we consider are the genus statistics. The genus
statistics have been attractive because it has a geometrical meaning
of the clustering as well as the cosmological significance. The 2D
genus statistic $G_2$ is defined in a two-dimensional plane $S$ in a
$d$-dimensional space, so that $d \geq 2$ is required. In this plane
$S$, there are contours corresponding to each threshold $\nu$. The 2D
genus statistic is defined by the number of contours surrounding
regions higher than the threshold value minus the number of contours
surrounding regions lower than the threshold value
\citep{adl81,col88,mel89,got90}. This definition is intuitive, but an
alternative, equivalent definition is more useful: first we set an
arbitrary, fixed direction on the plane $S$. Then there are maxima and
minima of contours according to that direction. These points are
classified into upcrossing points and downcrossing points with respect
to that chosen direction. The number of those points are used to
define the 2D genus statistic as the following way:
\begin{eqnarray}
&&
   G_2 =
   \frac12
   \left[
      (\mbox{\# of upcrossing minima})
     - (\mbox{\# of upcrossing maxima})
   \right.
\nonumber\\
&&\qquad\qquad
   \left.
     -\, (\mbox{\# of downcrossing minima})
     +  (\mbox{\# of downcrossing maxima})
   \right],
\label{eq2-8}
\end{eqnarray}
per unit area of the surface. According to this definition, the
explicit expression of the 2D genus statistic is given by
\begin{eqnarray}
   G_2(\nu) = 
   - \frac12
   \left\langle
      \delta(\alpha - \nu) \delta(\eta_1)|\eta_2|\zeta_{11}
   \right\rangle.
\label{eq2-9}
\end{eqnarray}

{}For this statistic, the indices $\mu_1, \mu_2 , \ldots$ only take
$0$, $1$, $2$ and $(11)$. Let the number of $0$ be $k$, $1$ be $l_1$,
$2$ be $l_2$, and $(11)$ be $m$. Then we need to calculate the
following quantity:
\begin{eqnarray}
&&
   \left\langle F_{,\mu_1\mu_2\cdots}\right\rangle_{\rm G} =
   R(k;l_1,l_2;m)
   \equiv
   -\frac12
   \left\langle
      \left(\frac{\partial}{\partial\alpha}\right)^k
      \left(\frac{\partial}{\partial\eta_1}\right)^{l_1}
      \left(\frac{\partial}{\partial\eta_2}\right)^{l_2}
      \left(\frac{\partial}{\partial\zeta_{11}}\right)^m
      \delta(\alpha - \nu) \delta(\eta_1)|\eta_2|\zeta_{11}
   \right\rangle_{\rm G}.
\label{eq2-10}
\end{eqnarray}
Since the second derivative $\zeta_{11}$ appears as a polynomial, we
just use the transform of equation (\ref{eq1-13-3-1}). Then, from
equations (\ref{eqa1-4a}), (\ref{eqa1-5a}) and (\ref{eqa1-5b}), the
above equation reduces to
\begin{eqnarray}
   R(k;l_1,l_2;m) =
    \frac{h_{l_1} h_{l_2-2}}{(2\pi)^{3/2}}
   \left(\frac{\sigma_1}{\sqrt{d}\sigma_0}\right)^{2-l_1-l_2-2m}
   e^{-\nu^2/2}
   \left[
      H_{k+1}(\nu) \delta_{m0} -
      H_{k}(\nu) \delta_{m1}
   \right].
\label{eq2-11}
\end{eqnarray}
Thus, from equation (\ref{eq1-17}),
\begin{eqnarray}
   \sum_{\mu,\nu,\lambda} \hMt_{\mu\nu\lambda} F_{\mu\nu\lambda}
   &=&
   - 2 S^{(1)}
   \left( \frac{\sigma_1}{\sqrt{d}\sigma_0} \right)^2
   \left[ 2 R(2;0,0;1) - R(1;2,0;0) - R(1;0,2;0) \right]
   -
   2 S^{(2)}
   \left( \frac{\sigma_1}{\sqrt{d}\sigma_0} \right)^4
   R(0;0;2,1)
\nonumber\\
   &=&
   \frac{1}{(2\pi)^{3/2}} 
   \left( \frac{\sigma_1}{\sqrt{d}\sigma_0} \right)^2
   e^{-\nu^2/2}
   \left[
      S^{(0)} H_4(\nu) + 4 S^{(1)} H_2(\nu) +
      2 S^{(2)}
   \right].
\label{eq2-12}
\end{eqnarray}
On the other hand, the Gaussian contribution is simply given by
\begin{eqnarray}
   \langle F \rangle_{\rm G} =
   R(0;0,0;0) =
   \frac{1}{(2\pi)^{3/2}}
   \left(\frac{\sigma_1}{\sqrt{d}\sigma_0}\right)^2
   e^{-\nu^2/2} H_1(\nu).
\label{eq2-13}
\end{eqnarray}
Thus, the perturbative expansion of equation (\ref{eq1-13}) up to
second order is finally given by
\begin{eqnarray}
&&
   G_2(\nu) =
   \frac{1}{(2\pi)^{3/2}}
   \left(\frac{\sigma_1}{\sqrt{d}\sigma_0}\right)^2
   e^{-\nu^2/2}
   \left\{
      H_1(\nu)
      +
      \left[
         \frac{S^{(0)}}{6} H_4(\nu) +
         \frac{2 S^{(1)}}{3} H_2(\nu) +
         \frac{S^{(2)}}{3}
      \right] \sigma_0 +
      {\cal O}({\sigma_0}^2)
   \right\}.
\label{eq2-14}
\end{eqnarray}

\subsection{3D Genus Statistic}
\label{sec3-4}

The second order formula for 3D genus statistic was already derived by
\citet{mat94}, but the detailed derivation was omitted. For
completeness, we revisit the same quantity from our general point of
view here. While the 2D genus statistic is defined by the number of
contour lines in 2D surface, the 3D genus statistic \citep{got86} is
defined by the number of contour surfaces and the number of handles in
3D space. Thus the 3D genus is defined only for cosmic fields of
$d=3$. The 3D genus statistic $G_3$ is defined by
\begin{eqnarray}
&&
   G_3 =
   \left[
      (\mbox{\# of handles of contours}) -
      (\mbox{\# of isolated contours})
   \right],
\label{eq2-15}
\end{eqnarray}
per unit volume of the 3D space. This quantity is mathematically
equivalent to $-1/2$ times Euler characteristic of the contour
surfaces, and thus is proportional to the total surface integral of
local curvature of contours from the Gauss-Bonnet theorem. Although
those definition is intuitive, an alternative, equivalent definition
is more useful as in the 2D genus case. We set an arbitrary direction
in the 3D space. Then there are maxima, minima and saddle points
according to that direction. From the number of these points, the 3D
genus is defined by
\begin{eqnarray}
&&
   G_3 =
   - \frac12
   \left[
      (\mbox{\# of maxima}) +
      (\mbox{\# of minima})
   -  (\mbox{\# of saddle points})
   \right],
\label{eq2-16}
\end{eqnarray}
per unit volume. According to this definition, the explicit expression
of the 3D genus statistic is given by \citep{dor70,adl81,bar86}
\begin{eqnarray}
   G_3(\nu) =
   - \frac12
   \left\langle
      \delta(\alpha - \nu)
      \delta(\eta_1) \delta(\eta_2) |\eta_3|
      \left(
         \zeta_{11}\zeta_{22} - {\zeta_{12}}^2
      \right)
   \right\rangle.
\label{eq2-17}
\end{eqnarray}

We need to calculate the following quantity:
\begin{eqnarray}
   \left\langle F_{,\mu_1\mu_2\cdots}\right\rangle_{\rm G} &=&
   R(k;l_1,l_2,l_3;m_{11},m_{22},m_{12})
\nonumber\\
   &\equiv&
   -\frac12
   \left\langle
      \left(\frac{\partial}{\partial\alpha}\right)^k
      \prod_{i=1}^3
      \left(\frac{\partial}{\partial\eta_i}\right)^{l_i}
      \prod_{i\leq j}^2
      \left(\frac{\partial}{\partial\zeta_{ij}}\right)^{m_{ij}}
      \delta(\alpha - \nu)
      \delta(\eta_1) \delta(\eta_2) |\eta_3|
      \left(
         \zeta_{11}\zeta_{22} - {\zeta_{12}}^2
      \right)
   \right\rangle_{\rm G}.
\label{eq2-18}
\end{eqnarray}
After applying the transform of equation (\ref{eq1-13-3-1}), the
Gaussian integration of $\widetilde{\zeta}_{ij}$ fixing $\alpha$ is
given by
\begin{eqnarray}
&&
   \left\langle
      \left.
      \prod_{i\leq j}^2
      \left(\frac{\partial}{\partial\zeta_{ij}}\right)^{m_{ij}}
      \left(
         \zeta_{11}\zeta_{22} - {\zeta_{12}}^2
      \right)
      \right| \alpha
   \right\rangle_{\rm G}
\nonumber\\
&&\qquad\qquad =
   \left(
      \frac{{\sigma_1}^2}{d\,{\sigma_0}^2}
   \right)^{2-\sum_{i\leq j} m_{ij}}
   \left[
      H_2(\alpha) J^{(2)}_0(\{m_{ij}\}) -
      H_1(\alpha) J^{(2)}_1(\{m_{ij}\}) +
      J^{(2)}_2(\{m_{ij}\})
   \right],
\label{eq2-19}
\end{eqnarray}
where $J^{(2)}_m$ is defined in Table~\ref{tab1}.
\begin{table}
\begin{center}
\caption{Definition of $J^{(2)}_m$. For other sets of
$\{m_{ij}\}$ not listed in this table, $J^{(2)}_m = 0$.
\label{tab1}}
\begin{tabular}{ccc|ccc}
\tableline\tableline
$m_{11}$ & $m_{22}$ & $m_{12}$ & $J^{(2)}_0$ & $J^{(2)}_1$ & $J^{(2)}_2$\\
\tableline
$0$ & $0$ & $0$ & $1$ & $0$ & $ 0$ \\ 
$1$ & $0$ & $0$ & $0$ & $1$ & $ 0$ \\ 
$0$ & $1$ & $0$ & $0$ & $1$ & $ 0$ \\ 
$1$ & $1$ & $0$ & $0$ & $0$ & $ 1$ \\ 
$0$ & $0$ & $2$ & $0$ & $0$ & $-2$ \\ 
\tableline
\end{tabular}
\end{center}
\end{table}
Then, from equations (\ref{eqa1-4a}), (\ref{eqa1-5a}) and
(\ref{eqa1-5b}), the equation (\ref{eq2-18}) reduces to
\begin{eqnarray}
&&   R(k;l_1,l_2,l_3;m_{11},m_{22},m_{12})
\nonumber\\
&&\qquad = 
   - \frac{1}{(2\pi)^2}
   \left(
      \frac{\sigma_1}{\sqrt{d}\,\sigma_0}
   \right)^{3 - \sum_{i=1}^3 l_i - 2 \sum_{i\leq j}^2 m_{ij}}
   h_{l_1} h_{l_2} h_{l_3-2}
   e^{-\nu^2/2}
\nonumber\\
&&\qquad\quad\times\,
   \left[
      J^{(2)}_0(\{m_{ij}\}) H_{k+2}(\nu) -
      J^{(2)}_1(\{m_{ij}\}) H_{k+1}(\nu) +
      J^{(2)}_2(\{m_{ij}\}) H_{k}(\nu)
   \right].
\label{eq2-20}
\end{eqnarray}
Since the 3D genus is only defined for $d \geq 3$ and our universe has
the spatial dimension 3, only $d=3$ is meaningful for actual cosmic
fields. Nevertheless, we preserve the general dimension $d$ for some
flavor of generality. Thus, from equation (\ref{eq1-17}),
\begin{eqnarray}
&&
   \sum_{\mu,\nu,\lambda} \hMt_{\mu\nu\lambda} F_{\mu\nu\lambda}
   =
   - \frac{1}{(2\pi)^2} 
   \left( \frac{\sigma_1}{\sqrt{d}\sigma_0} \right)^3
   e^{-\nu^2/2}
   \left[
      S^{(0)} H_5(\nu) + 6 S^{(1)} H_3(\nu) +
      6 S^{(2)} H_1(\nu)
   \right].
\label{eq2-21}
\end{eqnarray}
The Gaussian contribution is given by \citep{dor70,ham86}
\begin{eqnarray}
   \langle F \rangle_{\rm G} =
   R(0;0,0,0;0,0,0) =
   - \frac{1}{(2\pi)^2}
   \left(\frac{\sigma_1}{\sqrt{d}\sigma_0}\right)^3
   e^{-\nu^2/2} H_2(\nu).
\label{eq2-22}
\end{eqnarray}
Thus, the perturbative expansion of equation (\ref{eq1-13}) up to
second order is finally given by
\begin{eqnarray}
&&
   G_3(\nu) =
   - \frac{1}{(2\pi)^2}
   \left(\frac{\sigma_1}{\sqrt{d}\sigma_0}\right)^3
   e^{-\nu^2/2}
   \left\{
      H_2(\nu)
      +
      \left[
         \frac{S^{(0)}}{6} H_5(\nu) +
         S^{(1)} H_3(\nu) +
         S^{(2)} H_1(\nu)
      \right] \sigma_0 +
      {\cal O}({\sigma_0}^2)
   \right\}.
\label{eq2-23}
\end{eqnarray}
The above equation with $d=3$ agrees with \citet{mat94}\footnote{The
notations $S$, $T$, and $U$ in \citet{mat94} are related to the
notations here by $S = S^{(0)}$, $T = 2S^{(1)}/3$, $U = S^{(2)}/3$.}

\subsection{2D Weighted Extrema Density}
\label{sec3-5}

Next, we consider the weighted extrema density above a threshold
$\nu$. The field extrema is defined to be points where all the
first-order spatial derivatives of the field $f$ vanish: $\partial
f/\partial x_i = 0$. The weight $\pm 1$ is associated to each extrema
according to the number of negative eigenvalues of spatial
second-order derivatives of the field. When the threshold is high
enough, the weighted extrema are approximately identified with the
field peaks.

Mathematically, this statistic is equivalent to the 2D genus
statistics, according to the Morse's theorem \citep{mor69,adl81}.
Therefore, we do not need to calculate separately to obtain the result
on this statistic. However, we present the derivation of the extrema
density as an alternative calculation to the 2D genus.

The weighted extrema in 2D field is given by
\begin{eqnarray}
   \rho_{\rm e2}(\nu) =
   \left\langle
   \theta(\alpha - \nu)
   \delta(\eta_1) \delta(\eta_2)
   \left(\zeta_{11}\zeta_{22} - {\zeta_{12}}^2\right)
   \right\rangle.
\label{eq2-24}
\end{eqnarray}
Following the similar calculation of previous examples, and using
equations in Appendix~\ref{app1}, we obtain
\begin{eqnarray}
   \left\langle F_{,\mu_1\mu_2\cdots}\right\rangle_{\rm G} &=&
   R(k;l_1,l_2;m_{11},m_{22},m_{12})
\nonumber\\
   &\equiv&
   -\frac12
   \left\langle
      \left(\frac{\partial}{\partial\alpha}\right)^k
      \prod_{i=1}^2
      \left(\frac{\partial}{\partial\eta_i}\right)^{l_i}
      \prod_{i\leq j}^2
      \left(\frac{\partial}{\partial\zeta_{ij}}\right)^{m_{ij}}
      \delta(\alpha - \nu) \delta(\eta_1) \delta(\eta_2)
      \left(\zeta_{11}\zeta_{22} - {\zeta_{12}}^2\right)
   \right\rangle_{\rm G}
\nonumber\\
   &=&
   \frac{1}{(2\pi)^{3/2}}
   \left(
      \frac{\sigma_1}{\sqrt{d}\,\sigma_0}
   \right)^{2 - \sum_{i=1}^2 l_i - 2 \sum_{i\leq j}^2 m_{ij}}
   h_{l_1} h_{l_2}
   e^{-\nu^2/2}
\nonumber\\
&&\qquad\quad\times\,
   \left[
      J^{(2)}_0(\{m_{ij}\}) H_{k+1}(\nu) -
      J^{(2)}_1(\{m_{ij}\}) H_{k}(\nu) +
      J^{(2)}_2(\{m_{ij}\}) H_{k-1}(\nu)
   \right],
\label{eq2-25}
\end{eqnarray}
and the perturbative expansion of equation (\ref{eq1-13}) up to second
order is finally given by
\begin{eqnarray}
&&
   \rho_{\rm e2}(\nu) =
   \frac{1}{(2\pi)^{3/2}}
   \left(\frac{\sigma_1}{\sqrt{d}\sigma_0}\right)^2
   e^{-\nu^2/2}
   \left\{
      H_1(\nu)
      +
      \left[
         \frac{S^{(0)}}{6} H_4(\nu) +
         \frac{2 S^{(1)}}{3} H_2(\nu) +
         \frac{S^{(2)}}{3}
      \right] \sigma_0 +
      {\cal O}({\sigma_0}^2)
   \right\}.
\label{eq2-26}
\end{eqnarray}
This result is in fact equivalent to the 2D genus statistics given by
equation (\ref{eq2-14}). 

\subsection{Minkowski Functionals}
\label{sec3-6}

Most of the Minkowski functionals are closely related to statistics
discussed above. In this subsection, we comprehensively describe the
exact relation between the Minkowski functionals $V^{(d)}_k$ of a
smoothed field and the statistical quantities considered above.

The Minkowski functional of $k=0$ is simply the volume fraction of the
excursion set $K$ which is defined by high-density regions above a
given threshold $\nu$:
\begin{eqnarray}
   V^{(d)}_0(\nu) = \frac{1}{V}
   \int_{K} dV.
\label{eq2-50-0}
\end{eqnarray}

The other functionals with $k=1,2,\ldots,d$ are defined by the
integral of the curvatures on isodensity surfaces of the threshold
$\nu$ \citep{sch97,sch98}. In 3-dimensions, $d=3$, they are evaluated
by a surface integration averaged over whole system of volume $V$
\citep{sch97}, i.e.,
\begin{eqnarray}
   V^{(3)}_k(\nu) = \frac{1}{V}
   \int_{\partial K} d^2 A(\bfx) v^{(3)}_k(\nu,\bfx),
\label{eq2-50}
\end{eqnarray}
where the local Minkowski functionals,
\begin{eqnarray}
&&
   v^{(3)}_1(\nu,\bfx) = \frac16,
\label{eq2-51a}\\
&&
   v^{(3)}_2(\nu,\bfx) = \frac{1}{6\pi}
   \left(\frac{1}{R_1} + \frac{1}{R_2}\right),
\label{eq2-51b}\\
&&
   v^{(3)}_3(\nu,\bfx) = \frac{1}{4\pi}\frac{1}{R_1 R_2},
\label{eq2-51c}
\end{eqnarray}
are defined by the principal curvatures $1/R_1$ and $1/R_2$ of the
surface oriented toward lower density values. In 2-dimensions,
$d=2$, the Minkowski functionals of $k=1,2$ are evaluated by a line
integration averaged over whole system of 2D volume (surface) $V$
\citep{sch98}, i.e.,
\begin{eqnarray}
   V^{(2)}_k(\nu) = \frac{1}{V}
   \int_{\partial K} dL(\bfx) v^{(2)}_k(\nu,\bfx),
\label{eq2-52}
\end{eqnarray}
where the local Minkowski functionals,
\begin{eqnarray}
&&
   v^{(2)}_1(\nu,\bfx) = \frac14,
\label{eq2-53a}\\
&&
   v^{(2)}_2(\nu,\bfx) = \frac{1}{2\pi} \frac{1}{R_1},
\label{eq2-53b}
\end{eqnarray}
are defined by the principal curvature $1/R_1$ of the line oriented
toward lower density values.

All the Minkowski functionals for a Gaussian random field are
analytically derived by \citet{tom86}:
\begin{eqnarray}
   V^{(d)}_k(\nu) = 
   \frac{1}{(2\pi)^{(k+1)/2}}
   \frac{\omega_d}{\omega_{d-k}\omega_k}
   \left(\frac{\sigma_1}{\sqrt{d}\sigma_0}\right)^k
   e^{-\nu^2/2}
   H_{k-1}(\nu),
\label{eq2-54}
\end{eqnarray}
where the factor $\omega_k = \pi^{k/2}/\Gamma(k/2 + 1)$ is the volume
of the unit ball in $k$ dimensions, so that $\omega_0 = 1$, $\omega_1
= 2$, $\omega_2 = \pi$, $\omega_3 = 4\pi/3$ \citep{sch97}.

It turns out that the Minkowski functionals in 2- and 3-dimensions are
identical to the statistics $N_1$, $G_2$ and $G_3$ in each dimensions,
except for normalization factors. In fact, from the Crofton's formula
\citep{cro1868}, the $k$-th Minkowski functional is given by
\begin{eqnarray}
   V^{(d)}_k =
   \frac{\omega_d}{\omega_{d-k}\omega_k}
   \int_{{\cal E}^{(d)}_k} d\mu_k(E) \chi^{(k)}(K \cap E).
\label{eq2-100}
\end{eqnarray}
In this formula for body $K$ in $d$ dimensions, we consider an
arbitrary $k$-dimensional hypersurface $E$ and calculate the Euler
characteristic $\chi^{(k)}$ of the intersection $K \cap E$ in $k$
dimensions. This quantity is integrated over the space ${\cal
E}^{(d)}_k$ of all conceivable hypersurfaces. The integration measure
$d\mu_k(E)$ is normalized to give $\int_{{\cal E}_k^{(d)}} d\mu_k(E) =
1$. From this formula, we can see the statistics $G_3$, $G_2$, $N_1$
are proportional to the Minkowski functionals of $V^{(d)}_3$,
$V^{(d)}_2$, $V^{(d)}_1$. In fact, $\chi^{(3)}$ is given by $-1$ times
the 3D genus (or, equivalently, $1/2$ times the Euler number of
boundaries, $\chi(\partial(K \cap E))$), $\chi^{(2)}$ is identical to
the 2D genus, and $\chi^{(1)}$ is just $1/2$ times the number of
level-crossing points \citep{adl81}. Thus, Minkowski functionals are
given by
\begin{eqnarray}
&&
   V^{(d)}_3(\nu) =
   - \frac{\omega_d}{\omega_{d-3}\omega_3} G_3(\nu),
\label{eq2-101a}\\
&&
   V^{(d)}_2(\nu) = \frac{\omega_d}{\omega_{d-2}\omega_2} G_2(\nu),
\label{eq2-101b}\\
&&
   V^{(d)}_1(\nu) = \frac{\omega_d}{2\ \omega_{d-1}\omega_1} N_1(\nu),
\label{eq2-101c}
\end{eqnarray}
where the boundary of the body $K$ is identified with the isodensity
contours of threshold $\nu$ and the statistics on right hand sides are
defined in $d$-dimensions. Therefore, we have already obtained the
weakly non-Gaussian expressions for Minkowski functionals, i.e., the
Minkowski functionals of $k=1,2,3$ are given by the above equations
and equations (\ref{eq2-7}), (\ref{eq2-14}), and (\ref{eq2-23}). The
Gaussian parts of the above equations exactly reproduce the Tomita's
formula (\ref{eq2-54}).

The remaining Minkowski functional is the volume functional
\begin{eqnarray}
   V^{(d)}_0(\nu) =
   \left\langle \theta(\nu - \alpha) \right\rangle.
\label{eq2-102}
\end{eqnarray}
In this case, from equation (\ref{eqa1-4a}) in Appendix~\ref{app1},
\begin{eqnarray}
   R(k) \equiv
   \left\langle
      \left(\frac{\partial}{\partial\alpha}\right)^k
      \theta(\alpha - \nu)
   \right\rangle_{\rm G} =
   \frac{e^{-\nu^2/2}}{\sqrt{2\pi}} H_{k-1}(\nu).
\label{eq2-103}
\end{eqnarray}
so that equation (\ref{eq1-17}) reduces to
\begin{eqnarray}
   V^{(d)}_0(\nu) =
   \frac12 {\rm erfc}\left(\frac{\nu}{\sqrt{2}}\right) +
   \frac{e^{-\nu^2/2}}{\sqrt{2\pi}} \frac{S^{(0)}}{6}
   H_2(\nu) \sigma_0 +
   {\cal O}({\sigma_0}^2).
\label{eq2-104}
\end{eqnarray}
The equivalent form can be obtained by integrating the Edgeworth
expansion of PDF, equation (\ref{eq2-0-2}). 

All the formulas of Minkowski functionals derived above for $0 \leq k
\leq 3$ fit into a single expression:
\begin{eqnarray}
   V^{(d)}_k(\nu) &=&
   \frac{1}{(2\pi)^{(k+1)/2}}
   \frac{\omega_d}{\omega_{d-k}\omega_k}
   \left(\frac{\sigma_1}{\sqrt{d}\sigma_0}\right)^k
   e^{-\nu^2/2}
\nonumber\\
   && \qquad\times
   \left\{
      H_{k-1}(\nu) +
      \left[
         \frac16 S^{(0)} H_{k+2}(\nu) +
         \frac{k}{3} S^{(1)} H_k(\nu) +
         \frac{k(k-1)}{6} S^{(2)} H_{k-2}(\nu)
      \right] \sigma_0 +
      {\cal O}({\sigma_0}^2)
   \right\}.
\label{eq2-60}
\end{eqnarray}

\subsection{Rescaling the Threshold Density by Volume Fractions}
\label{sec3-7}

The density threshold $\nu$ so far is simply defined so that
isodensity surface is identified by $f=\nu\sigma_0$. However, the
horizontal shift of the nonlinear genus curve etc.~is considerably
attributed to the nonlinear shift of probability distribution of the
density field \citep[e.g.,][]{got87,mat96c}. In order to cancel the
latter shift, the threshold $\tnu$ is defined so that the volume
fraction $f_V$ on the high-density side of the isodensity surface
equals to
\begin{eqnarray}
   f_V = \frac{1}{\sqrt{2\pi}}
   \int_\tnu^\infty dt e^{-t^2/2}.
\label{eq2-105}
\end{eqnarray}
In fact, most of the work on genus analysis uses the genus curve
plotted against the volume-fraction threshold $\tnu$. Recently,
\citet{set00} re-expressed the weakly non-Gaussian formula of genus
curve \citep{mat94} in terms of $\tnu$, using perturbative expansion
of the probability distribution function of the density field
\citep{jus95}. We follow this method to re-express the weakly
non-Gaussian formulas of various statistical quantities derived above
in terms of the volume-fraction threshold $\tnu$.

The relation of $\nu$ and $\tnu$ of weakly non-Gaussian field is
simply given by equating the two equations (\ref{eq2-104}) and
(\ref{eq2-105}). Up to first order in $\sigma_0$, the relation reduces
to
\begin{eqnarray}
   \nu = \tnu + \frac{S^{(0)}}{6} H_2(\tnu) \sigma_0
   + {\cal O}({\sigma_0}^2).
\label{eq2-106}
\end{eqnarray}
It is straightforward to rewrite the various analytical formulas we
derived above. The results for level-crossing statistic, 2D and 3D
genus are respectively given by
\begin{eqnarray}
   N_1(\tnu) &=&
   \frac{1}{\pi} \frac{\sigma_1}{\sqrt{d}\sigma_0} e^{-\tnu^2/2}
   \left\{
      1 +
      \left[
         \frac13\left(S^{(1)} - S^{(0)}\right) H_1(\tnu)
      \right] \sigma_0 +
      {\cal O}({\sigma_0}^2)
   \right\},
\label{eq2-207a}\\
   G_2(\tnu) &=& \rho_{\rm e2}(\tnu) =
   \frac{1}{(2\pi)^{3/2}}
   \left(\frac{\sigma_1}{\sqrt{d}\sigma_0}\right)^2
   e^{-\tnu^2/2}
\nonumber\\
&&\qquad\qquad\quad\times\,
   \left\{
      H_1(\tnu)
      +
      \left[
         \frac23 \left(S^{(1)} - S^{(0)}\right) H_2(\tnu) +
         \frac13 \left(S^{(2)} - S^{(0)}\right)
      \right] \sigma_0 +
      {\cal O}({\sigma_0}^2)
   \right\},
\label{eq2-207b}\\
   G_3(\tnu) &=&
   - \frac{1}{(2\pi)^2}
   \left(\frac{\sigma_1}{\sqrt{d}\sigma_0}\right)^3
   e^{-\tnu^2/2}
\nonumber\\
&&\qquad\qquad\times\,
   \left\{
      H_2(\tnu)
      +
      \left[
         \left(S^{(1)} - S^{(0)}\right) H_3(\tnu) +
         \left(S^{(2)} - S^{(0)}\right) H_1(\tnu)
      \right] \sigma_0 +
      {\cal O}({\sigma_0}^2)
   \right\},
\label{eq2-207c}
\end{eqnarray}
The results for Minkowski functionals of $k=1,2,3$ are again given by
these equations and equations (\ref{eq2-101a})--(\ref{eq2-101c}). All
the Minkowski functionals for $0 \leq k \leq 3$ fit into a single
expression:
\begin{eqnarray}
   V^{(d)}_k(\tnu) &=& 
   \frac{1}{(2\pi)^{(k+1)/2}}
   \frac{\omega_d}{\omega_{d-k}\omega_k}
   \left(\frac{\sigma_1}{\sqrt{d}\sigma_0}\right)^k
   e^{-\tnu^2/2}
\nonumber\\
   && \qquad\times
   \left\{
      H_{k-1}(\tnu) +
      \left[
         \frac{k}{3} \left(S^{(1)} - S^{(0)}\right) H_k(\tnu) +
         \frac{k(k-1)}{6} \left(S^{(2)} - S^{(0)}\right) H_{k-2}(\tnu)
      \right] \sigma_0 +
      {\cal O}({\sigma_0}^2)
   \right\}.
\label{eq2-208}
\end{eqnarray}
Remarkably, the highest-order Hermite polynomial in the non-Gaussian
correction terms vanishes in each case. In addition, the skewness
parameters only appear as combinations of the form, $S^{(a)} -
S^{(0)}$, which makes the result simpler compared with the original
form with the direct threshold $\nu$. As we will see in the following
sections, the numerical values of $S^{(a)}, (a=0,1,2)$ are quite
close, or even identical in some special models. This means that the
non-Gaussian corrections of the above statistics are smaller with the
rescaled threshold $\tnu$ than with the original threshold $\nu$. This
is one of the central results in this paper. This tendency is in
agreement with the analyses of numerical simulations.

\setcounter{equation}{0}
\section{SKEWNESS PARAMETERS FOR SMOOTHED FIELDS}
\label{sec4}

We need to know the skewness parameters, $S^{(a)}$ for the evaluation
of the second-order perturbative terms of equation (\ref{eq1-17}).
These quantities can be calculated by usual perturbation theory, once
we specify the cosmic field, $f$. In each example of the previous
section, the quantity $S^{(2)}_2$ does not appear so that we evaluate
$S^{(0)}$, $S^{(1)}$, and $S^{(2)}$ in this section. The other kinds
of skewness parameters like $S^{(2)}_2$ can be similarly evaluated
without difficulty.

\subsection{Hierarchical Model}
\label{sec4-1}

Before we explore actual cosmic fields, we consider a simple,
phenomenological statistical model, i.e., the hierarchical model of
higher order correlation functions \citep[e.g.,][]{pee80}. In this
model, $N$-point correlation function is a sum of $N-1$ products of
the two-point correlation function. Specifically, the three-point
correlation function is given by
\begin{eqnarray}
&&
   \left\langle
      f(\bfx_1) f(\bfx_2) f(\bfx_3)
   \right\rangle =
   Q 
   \left[
      \langle f(\bfx_1) f(\bfx_2) \rangle
      \langle f(\bfx_2) f(\bfx_3) \rangle +
      \langle f(\bfx_2) f(\bfx_3) \rangle
      \langle f(\bfx_3) f(\bfx_1) \rangle
   \right.
\nonumber\\
&&\qquad\qquad\qquad\qquad\qquad\qquad\qquad\qquad
   \left. +\,
      \langle f(\bfx_3) f(\bfx_1) \rangle
      \langle f(\bfx_1) f(\bfx_2) \rangle
   \right].
\label{eq4-1}
\end{eqnarray}
We assume the field $f$ in the above equation is already smoothed. In
this case, skewness parameters, equations
(\ref{eq1-15a})--(\ref{eq1-15d}) are given by straightforward
calculations \citep{mat94}:
\begin{eqnarray}
   S^{(0)} = S^{(1)} = S^{(2)} = 3 Q.
\label{eq4-2}
\end{eqnarray}
These values depend on a hierarchical amplitude, $Q$, which is a free
parameter of this model. The relative amplitudes among $S^{(a)}$ are
not freely adjusted in the above equation. In the case of
volume-fraction threshold, the first non-Gaussian correction of the
various statistics considered in the previous section is absent, since
they depend only on $S^{(a)} - S^{(0)}$.

\subsection{3D Density Field}
\label{sec4-2}

Next, we consider the three-dimensional density field. The skewness
parameters of this field in perturbation theory are already calculated
by \citet{mat94} and \citet{mat96b}, using Fourier transforms of the
field. We comprehensively review this calculation here for
completeness. There is an alternative way to calculate the skewness
not depending on Fourier transforms \citep{buc99}.

The cosmic field $f$ is identified with the 3D density contrast,
$\rho/\bar{\rho}-1$, where $\rho$ is the density field. The dimension
this field is defined in is three, $d=3$. The Fourier transform of the
field is useful in the following:
\begin{eqnarray}
   \widetilde{f}(\bfk) =
   \int d^3x e^{-i\sbk\cdot\sbx} f(\bfx).
\label{eq4-3}
\end{eqnarray}
In this notation, two- and three-point correlations in Fourier space
have the forms,
\begin{eqnarray}
&&
   \left\langle 
      \widetilde{f}(\bfk_1)\widetilde{f}(\bfk_2)
   \right\rangle =
   (2\pi)^2 \delta^3(\bfk_1 + \bfk_2)
   P(k_1),
\label{eq4-4a}\\
&&
   \left\langle 
      \widetilde{f}(\bfk_1)\widetilde{f}(\bfk_2)
      \widetilde{f}(\bfk_3)
   \right\rangle =
   (2\pi)^3 \delta^3(\bfk_1 + \bfk_2 + \bfk_3)
   B(k_1, k_2, k_3).
\label{eq4-4b}
\end{eqnarray}
The above forms are the consequence of the statistical homogeneity of
the space, where the functions $P$ and $B$ are the power spectrum and
the bispectrum, respectively. Thus, the variance and its variants
[Eqs.~(\ref{eq1-2}), (\ref{eq1-13-3a}), and (\ref{eq1-13-3b})], in
their Fourier representation, are given by
\begin{eqnarray}
   {\sigma_j}^2 =
   \int \frac{k^2dk}{2\pi^2} k^{2j} P(k),
\label{eq4-5}
\end{eqnarray}
and the skewness parameters of equations
(\ref{eq1-15a})--(\ref{eq1-15d}) are given by
\begin{eqnarray}
&&
   S^{(0)} = \frac{1}{{\sigma_0}^4}
   \int \frac{d^3k_1}{(2\pi)^3} \frac{d^3k_2}{(2\pi)^3}
   B(k_1, k_2, |\bfk_1 + \bfk_2|),
\label{eq4-6a}\\
&&
   S^{(1)} = \frac{3}{{4 \sigma_0}^2 {\sigma_1}^2}
   \int \frac{d^3k_1}{(2\pi)^3} \frac{d^3k_2}{(2\pi)^3}
   |\bfk_1 + \bfk_2|^2
   B(k_1, k_2, |\bfk_1 + \bfk_2|),
\label{eq4-6b}\\
&&
   S^{(2)} = \frac{9}{{4 \sigma_1}^4}
   \int \frac{d^3k_1}{(2\pi)^3} \frac{d^3k_2}{(2\pi)^3}
   (\bfk_1\cdot\bfk_2)|\bfk_1 + \bfk_2|^2
   B(k_1, k_2, |\bfk_1 + \bfk_2|).
\label{eq4-6c}
\end{eqnarray}

For initial random Gaussian density field, the second order
perturbation theory predicts the power spectrum and the bispectrum as
follows \citep[e.g.,][]{pee80,fry84,bou92,ber94a}:
\begin{eqnarray}
&&   P(k) = \PL(k) W^2(kR) + {\cal O}({\sigma_0}^4),
\label{eq4-7a}\\
&&   B(k_1, k_2, k_3)
   =
   \left[
      1 + E + 
      \left(\frac{k_2}{k_1} + \frac{k_1}{k_2}\right)
      \frac{\bfk_1\cdot\bfk_2}{k_1 k_2} +
      (1 - E)
      \left(\frac{\bfk_1\cdot\bfk_2}{k_1 k_2}\right)^2
   \right]
   \PL(k_1) \PL(k_2) W(k_1R) W(k_2R) W(k_3R)
\nonumber\\
&&\qquad\qquad\qquad\quad +\,
   {\rm cyc.(1,2,3)}
   + {\cal O}({\sigma_0}^6),
\label{eq4-7b}
\end{eqnarray}
where $\PL(k)$ is the linear power spectrum, and $E$ is a weak
function of cosmology \citep{ber94a,ber95a}. The field smoothing
corresponds to the multiplication of the window function $W(kR)$ which
is the three-dimensional Fourier transform of the smoothing function
$W_R$. It is a good approximation to use the value of $E$ for an
Einstein-de Sitter universe, $E = 3/7$, in most cases. The explicit
form of the function $E$ in terms of cosmological parameters
$\Omega_0$ and $\lambda_0$ is given by \citet{mat95b}, which is
accurately fitted by
\begin{eqnarray}
   E \approx \frac37 {\Omega_0}^{-1/30} -
   \frac{\lambda_0}{80}
   \left(1 - \frac32 \lambda_0 \log_{10}\Omega_0\right).
\label{eq2-21-1}
\end{eqnarray}
The perturbation theory is considered to be an expansion by a
parameter $\sigma_0$. In this respect, the power spectrum $P$ and the
bispectrum $B$ is of order ${\sigma_0}^2$ and ${\sigma_0}^4$,
respectively.

Substituting equation (\ref{eq4-7a}) into equation (\ref{eq4-5}), we
obtain
\begin{eqnarray}
   {\sigma_j}^2 (R) = 
   \int \frac{k^2dk}{2\pi^2} k^{2j} \PL(k) W^2(kR),
\label{eq4-7-1}
\end{eqnarray}
up to the lowest order. Similarly, substituting equation
(\ref{eq4-7b}) into equations (\ref{eq4-6a})--(\ref{eq4-6c}), and
introducing new integration variables, $l_1 \equiv |\bfk_1|R$, $l_2
\equiv |\bfk_2|R$, and $\mu = \bfk_1\cdot\bfk_2/(k_1 k_2)$, we
obtain
\begin{eqnarray}
&&
   S^{(a)}(R) = \frac{1}{{\sigma_0}^4}
   \left(\frac{\sigma_0}{\sigma_1 R}\right)^{2a}
   \int \frac{{l_1}^2dl_1}{2\pi^2 R^3} \frac{{l_2}^2dl_2}{2\pi^2 R^3}
   \PL\left(\frac{l_1}{R}\right) \PL\left(\frac{l_2}{R}\right)
   W^2(l_1) W^2(l_2)
   \widetilde{S}^{(a)}(l_1,l_2),
\label{eq4-8}
\end{eqnarray}
where $a = 0,1,2$ and
\begin{eqnarray}
&&
   \widetilde{S}^{(a)}(l_1, l_2) =
   \frac{3}{4}
   \int_{-1}^1 d\mu
   \left[
      1 + E +
      \left(\frac{l_2}{l_1} + \frac{l_1}{l_2} \right) \mu +
      (1 - E) \mu^2
   \right]
   \frac{W\left(\sqrt{{l_1}^2 + {l_2}^2 + 2 l_1 l_2 \mu} \right)}
      {W(l_1) W(l_2)}
\nonumber\\
&&\qquad\qquad\qquad \times\,
   \left\{
   \begin{array}{ll}
      \displaystyle
      2, & a=0,\\
      {l_1}^2 + {l_2}^2 + l_1 l_2 \mu, & a=1,\\
      3\, {l_1}^2 {l_2}^2 \left( 1 - \mu^2 \right), & a=2,
   \end{array}
   \right.
\label{eq4-9}
\end{eqnarray}
So far the smoothing function is arbitrary. For a general smoothing
function, the above equations can be numerically integrated to obtain
the skewness parameters for each model of the power spectrum. {}For
some smoothing functions, further analytical reductions of the above
equations are possible. As a popular example, we consider the Gaussian
smoothing, $W(l) = \exp(-l^2/2)$, which is frequently adopted for
practical purposes. We follow the similar technique of \citet{lok95},
in which they derived the skewness of the density field with Gaussian
smoothing.

For the Gaussian smoothing,
\begin{eqnarray}
   \frac{W\left(\sqrt{{l_1}^2 + {l_2}^2 + 2 l_1 l_2 \mu} \right)}
      {W(l_1) W(l_2)} =
   e^{- l_1 l_2 \mu}.
\label{eq4-10}
\end{eqnarray}
In this case, the following formula of the modified Bessel function
$I_\nu(z)$ is useful:
\begin{eqnarray}
   \int_{-1}^1 d\mu P_l(\mu) e^{-\mu z} =
   (-1)^l \sqrt{\frac{2\pi}{z}} I_{l+1/2}(z),
\label{eq4-11}
\end{eqnarray}
where $P_l$ is the $l$-th Legendre polynomial. From this formula, the
angular integration of $\mu$ in equation (\ref{eq4-9}) can be
analytically performed and the result is
\begin{eqnarray}
&&
   \widetilde{S}^{(0)} =
   \sqrt{\frac{2\pi}{l_1l_2}}
   \left[
      (2 + E) I_{1/2}(l_1l_2) -
      \frac32 \left(\frac{l_2}{l_1} + \frac{l_1}{l_2}\right)
      I_{3/2}(l_1l_2) +
      (1 - E) I_{5/2}(l_1l_2)
   \right],
\label{eq4-12a}\\
&&
   \widetilde{S}^{(1)} =
   \sqrt{2\pi l_1l_2}
   \left\{
      \frac{5 + 2E}{4}
      \left(\frac{l_2}{l_1} + \frac{l_1}{l_2}\right)
      I_{1/2}(l_1l_2)
   -
      \left[
         \frac{3(9 + E)}{10} + 
         \frac{{l_2}^2}{{l_1}^2} + \frac{{l_1}^2}{{l_2}^2}
      \right]
      I_{3/2}(l_1l_2) 
   \right.
\nonumber\\
&&\qquad\qquad\qquad\qquad\qquad\qquad +\,
   \left.
      \frac{2 - E}{2}
      \left(\frac{l_2}{l_1} + \frac{l_1}{l_2}\right)
      I_{5/2}(l_1l_2)
   -
      \frac{3(1 - E)}{10}
      I_{7/2}(l_1 l_2)
   \right\},
\label{eq4-12b}\\
&&
   \widetilde{S}^{(2)} =
   \sqrt{2\pi}(l_1l_2)^{3/2}
   \left[
      \frac{3(3 + 2E)}{5}
      I_{1/2}(l_1l_2) -
      \frac{9}{10} \left(\frac{l_2}{l_1} + \frac{l_1}{l_2}\right)
      I_{3/2}(l_1l_2)
   \right.
\nonumber\\
&&\qquad\qquad\qquad\qquad\qquad -\,
   \left.
      \frac{3(3 + 4E)}{7}
      I_{5/2}(l_1l_2) +
      \frac{9}{10} \left(\frac{l_2}{l_1} + \frac{l_1}{l_2}\right)
      I_{7/2}(l_1l_2) -
      \frac{18(1 - E)}{35}
      I_{9/2}(l_1l_2)
   \right].
\label{eq4-12c}
\end{eqnarray}
At this point, it is useful to define the following quantity:
\begin{eqnarray}
&&
   S^{\alpha\beta}_m(R) \equiv
   \frac{\sqrt{2\pi}}{{\sigma_0}^4}
   \left(
      \frac{\sigma_0}{\sigma_1 R}
   \right)^{\alpha+\beta-2}
   \int \frac{{l_1}^2dl_1}{2\pi^2 R^3}
   \frac{{l_2}^2dl_2}{2\pi^2 R^3}
   \PL\left(\frac{l_1}{R}\right) \PL\left(\frac{l_2}{R}\right)
   e^{-{l_1}^2-{l_2}^2}
   {l_1}^{\alpha-3/2} {l_2}^{\beta-3/2}
   I_{m+1/2}(l_1l_2).
\label{eq4-13}
\end{eqnarray}
In the above equation, variance parameters $\sigma_0$, $\sigma_1$ are
given by equation (\ref{eq4-7-1}). The nonlinear correction for
$\sigma_j$ is not needed because our estimate of $S^{(a)}$ is only
lowest order in $\sigma_0$. When the higher order corrections, e.g.,
third-order perturbation corrections, are estimated, one should be
sure that all the necessary nonlinear corrections are properly taken
into account. With this quantity, the skewness parameters are given by
\begin{eqnarray}
&&
   S^{(0)}(R) =
   (2 + E) S^{11}_0 - 3 S^{02}_1 + (1 - E) S^{11}_2,
\label{eq4-14a}\\
&&
   S^{(1)}(R) =
   \frac32
   \left[
      \frac{5 + 2E}{3} S^{13}_0 -
      \frac{9 + E}{5} S^{22}_1 -
      S^{04}_1 +
      \frac{2(2-E)}{3} S^{13}_2 -
      \frac{1-E}{5} S^{22}_3
   \right],
\label{eq4-14b}\\
&&
   S^{(2)}(R) =
   9 
   \left[
      \frac{3 + 2E}{15} S^{33}_0 -
      \frac15 S^{24}_1 -
      \frac{3 + 4E}{21} S^{33}_2 +
      \frac15 S^{24}_3 -
      \frac{2(1-E)}{35} S^{33}_4
   \right].
\label{eq4-14c}
\end{eqnarray}
Thus the lowest order estimates of skewness parameters are given by
the above equations (\ref{eq4-13})--(\ref{eq4-14c}). For each given
power spectrum, the integration of equation (\ref{eq4-13}) is
straightforward.

The resulting skewness parameters are independent on the amplitude of
the power spectrum. When the power spectrum is given by a CDM-like
model, $\PL(k) \propto k {T_{\rm CDM}}^2 (k/\Gamma)$, where
\begin{eqnarray}
   T_{\rm CDM}(p) =
   \frac{\ln (1 + 2.34 p)}{2.34 p}
   \left[
      1 + 3.89 p + (16.1 p)^2 + (5.46 p)^3 + (6.71 p)^4
   \right]^{-1/4},
   \label{eq4-15}
\end{eqnarray}
is the CDM-like transfer function fitted by \citet{bar86}, and
$\Gamma$ is the shape parameter of this model, then the skewness
parameters are functions of $\Gamma R$. In Table~\ref{tab2}, we give
the values of skewness parameters $S^{(a)}$ for CDM-like models for
several values of $\Gamma R$.
\begin{table}
\begin{center}
\caption{3D skewness parameters of Gaussian-smoothed density
field for CDM-like models, which are the functions of the product of
shape parameter $\Gamma$ and smoothing length $R$. The parameter $E$
is set as $E=3/7$.
\label{tab2}}
\begin{tabular}{c|cccccc} \tableline\tableline
$\Gamma R$ & $1.0$ & $2.0$ & $4.0$ & $8.0$ &
$16.0$ & $32.0$\\
\tableline
$S^{(0)}$ & $3.678$ & $3.500$ & $3.332$ & $3.201$ &
$3.115$ & $3.065$\\
$S^{(1)}$ & $3.757$ & $3.566$ & $3.377$ & $3.228$ &
$3.129$ & $3.072$\\
$S^{(2)}$ & $3.657$ & $3.662$ & $3.701$ & $3.783$ &
$3.903$ & $4.046$\\
\tableline
\end{tabular}
\end{center}
\end{table}
In this table, the value of $E$ is approximated by $3/7$. Since the
skewness parameters are weak functions of $\Gamma R$ as seen from the
table, one can interpolate the values in this table to obtain the
values of arbitrary scales for practical purposes.

When the power spectrum is given by a power-law form,
\begin{eqnarray}
   \PL(k) = A k^\nspec,
\label{eq4-16}
\end{eqnarray}
the integration of equation (\ref{eq4-13}) can be analytically
performed. First, the simple Gaussian integration gives
\begin{eqnarray}
   {\sigma_j}^2 =
   \frac{A}{4\pi^2 R^{\nspec+2j+3}}
   \Gamma\left(\frac{\nspec+2j+3}{2}\right).
\label{eq4-17}
\end{eqnarray}
Second, we expand the modified Bessel function as
\begin{eqnarray}
   I_\nu(z) = \left(\frac{z}{2}\right)^\nu
   \sum_{r=0}^\infty
   \frac{(z/2)^{2r}}{r! \Gamma(\nu+r+1)}.
\label{eq4-18}
\end{eqnarray}
Then the equation (\ref{eq4-13}) reduces to
\begin{eqnarray}
&&
   S^{\alpha\beta}_m =
   \frac{2}{(2m+1)!!}
   \left(\frac{\nspec+3}{2}\right)^{1-(\alpha+\beta)/2}
   \left(\frac{\nspec+3}{2}\right)_{(\alpha+m-1)/2}
   \left(\frac{\nspec+3}{2}\right)_{(\beta+m-1)/2}
\nonumber\\
&&\qquad\qquad\qquad\qquad\qquad \times\,
   F\left(
      \frac{\nspec+\alpha+m+2}{2}, \frac{\nspec+\beta+m+2}{2},
      m + \frac32; \frac14
   \right),
\label{eq4-19}
\end{eqnarray}
where $(\alpha)_n = \Gamma(\alpha + n)/\Gamma(\alpha) =
\alpha(\alpha+1)\cdots(\alpha+n-1)$, and $F$
is the Gauss hypergeometric function:
\begin{eqnarray}
   F(\alpha,\beta,\gamma;z) =
   \sum_{r=0}^\infty
   \frac{(\alpha)_r (\beta)_r}{(\gamma)_r}
   \frac{z^r}{r!}.
\label{eq4-20}
\end{eqnarray}
The equations (\ref{eq4-14a})--(\ref{eq4-14c}) and equation
(\ref{eq4-19}) give the skewness parameters $S^{(a)}$. The following
recursion relations for hypergeometric function
\begin{eqnarray}
&&
   \alpha F(\alpha+1,\beta,\gamma+1;z) =
   \gamma F(\alpha,\beta,\gamma;z) + 
   (\alpha-\gamma) F(\alpha,\beta,\gamma+1;z),
\label{eq4-21a}\\
&&
   \alpha\beta F(\alpha+1,\beta+1,\gamma+1;z) =
   \frac{\gamma(\gamma-1)}{z}
   \left[
      F(\alpha,\beta,\gamma-1;z) - F(\alpha,\beta,\gamma;z)
   \right],
\label{eq4-21b}\\
&&
   (\gamma-\alpha)(\gamma-\beta)z
   F(\alpha,\beta,\gamma+1;z) +
   \gamma[(\alpha+\beta-2\gamma+1)z+\gamma-1]
   F(\alpha,\beta,\gamma;z)
\nonumber\\
&&\qquad\qquad\qquad \, +
   \gamma(\gamma-1)(z-1)
   F(\alpha,\beta,\gamma-1;z) = 0.
\label{eq4-21c}
\end{eqnarray}
simplify the result:
\begin{eqnarray}
&&
   S^{(0)}(\nspec) =
   S^{(1)}(\nspec)
   =
   3 F\left(\frac{\nspec+3}{2},\frac{\nspec+3}{2},\frac32;\frac14 \right)
   - (\nspec + 2 - 2E) 
   F\left(\frac{\nspec+3}{2},\frac{\nspec+3}{2}, \frac52;\frac14 \right),
\label{eq4-22a}\\
&& 
   S^{(2)}(\nspec) =
   3 F\left(\frac{\nspec+5}{2},\frac{\nspec+5}{2},\frac52;\frac14\right)
   - \frac35 (\nspec + 4 - 4E)
   F\left(\frac{\nspec+5}{2},\frac{\nspec+5}{2},\frac72;\frac14\right).
\label{eq4-22c}
\end{eqnarray}
In this power-law case, skewness parameters do not depend on scales
$R$ but only on power-law index, $\nspec$. Incidentally,
$S^{(0)}(\nspec)$ and $S^{(1)}(\nspec)$ are identical. This is just
the coincidence and is not generally the case when the power spectrum
is not given by power-law. Several numerical values are shown
in Table~\ref{tab3}, where $E = 3/7$ is assumed.
\begin{table}
\begin{center}
\caption{3D skewness parameters of Gaussian-smoothed density
field for power-law models, which are the functions of the spectral
index $n$.
\label{tab3}}
\begin{tabular}{c|ccccccccc} \tableline\tableline
\hfil
$n$ & $-3.0$ & $-2.5$ & $-2.0$ & $-1.5$ & $-1.0$ & $-0.5$ & $0.0$ &
$0.5$ & $1.0$\\
\tableline
$S^{(0,1)}$ &
$4.857$ & $4.400$ & $4.022$ & $3.714$ & $3.468$ & $3.280$ &
$3.144$ & $3.061$ & $3.029$\\
$S^{(2)}$ &
$3.815$ & $3.720$ & $3.665$ & $3.652$ & $3.680$ & $3.754$ &
$3.877$ & $4.054$ & $4.294$\\
\tableline
\end{tabular}
\end{center}
\end{table}

\subsection{3D Velocity Field}
\label{sec4-3}

Next, we consider the 3D velocity field as the cosmic field $f$. Since
the rotational components of the velocity field are decaying modes of
gravitational evolution in perturbation theory, we only consider
the rotation-free component. The cosmic field $f$ is identified with
the dimensionless scalar field,
\begin{eqnarray}
   f(\bfx) = \frac{1}{H}\nabla\cdot \bfv(\bfx),
\label{eq4-23}
\end{eqnarray}
where $H$ is the Hubble parameter. One can also consider other
quantities like radial component of the velocity field, $V =
\bfn\cdot\bfv$, where $\bfn$ is the line-of-sight normal vector. Those
quantities are more complicated than a simple divergence. We
illustrate only the simplest case in this paper. The second order
perturbation theory predicts power spectrum and bispectrum of the
velocity field as follows \citep[e.g.,][]{ber94a}:
\begin{eqnarray}
&&
   P(k) =
   {\logd}^2
   \PL(k) W^2(kR) + {\cal O}({\sigma_0}^4),
\label{eq4-24a}\\
&&
   B(k_1, k_2, k_3)
   =
   - {\logd}^3
   \left[
      1 + E_v + 
      \left(\frac{k_2}{k_1} + \frac{k_1}{k_2}\right)
      \frac{\bfk_1\cdot\bfk_2}{k_1 k_2} +
      (1 - E_v)
      \left(\frac{\bfk_1\cdot\bfk_2}{k_1 k_2}\right)^2
   \right]
   \PL(k_1) \PL(k_2) W(k_1R) W(k_2R) W(k_3R)
\nonumber\\
&&\qquad\qquad\qquad\quad +\,
   {\rm cyc.(1,2,3)}
   + {\cal O}({\sigma_0}^6).
\label{eq4-24b}
\end{eqnarray}
In the above equation, the factor $\logd$ is a logarithmic derivative
of the growth factor, given by
\begin{eqnarray}
   \logd(\Omega_0,\lambda_0) = \left.\frac{d\ln D}{d\ln a}\right|_0
   \approx
   {\Omega_0}^{4/7}
   + \frac{\lambda_0}{70}\left(1 + \frac{\Omega_0}{2}\right),
\label{eq4-25}
\end{eqnarray}
\citep{lig90,lah91} and $D$ is the linear growth factor and $a$ is the
expansion factor. The logarithmic derivative is evaluated at present.
The factor $E_v$ is a weak function of cosmology. It is a good
approximation to use the value for an Einstein-de Sitter universe,
$E_v = -1/7$. The explicit form of the function $E_v$ in terms of
$\Omega$ and $\lambda$ is given by \citet{mat95b}, which is accurately
fitted by
\begin{eqnarray}
   \frac{E_v + 1}{2} \approx \frac37 {\Omega_0}^{-11/200} -
   \frac{\lambda_0}{70}
   \left(1 - \frac73 \lambda_0 \log_{10}\Omega_0\right).
\label{eq2-25-1}
\end{eqnarray}

The similarity of the equations (\ref{eq4-24a}) and (\ref{eq4-24b})
for velocity field to the equations (\ref{eq4-7a}), (\ref{eq4-7b}) for
density field is obvious. We can easily see the skewness parameters
for the velocity field are obtained by similar equations as
(\ref{eq4-14a})--(\ref{eq4-14c}):
\begin{eqnarray}
&&
   S^{(0)}(R) =
   - \frac{1}{\logd}
   \left[
      (2 + E_v) S^{11}_0 - 3 S^{02}_1 + (1 - E_v) S^{11}_2
   \right],
\label{eq4-26a}\\
&&
   S^{(1)}(R) =
   - \frac{3}{2 \logd}
   \left[
      \frac{5 + 2E_v}{3} S^{13}_0 -
      \frac{9 + E_v}{5} S^{22}_1 -
      S^{04}_1 +
      \frac{2(2-E_v)}{3} S^{13}_2 -
      \frac{1-E_v}{5} S^{22}_3
   \right],
\label{eq4-26b}\\
&&
   S^{(2)}(R) =
   - \frac{9}{\logd}
   \left[
      \frac{3 + 2E_v}{15} S^{33}_0 -
      \frac15 S^{24}_1 -
      \frac{3 + 4E_v}{21} S^{33}_2 +
      \frac15 S^{24}_3 -
      \frac{2(1-E_v)}{35} S^{33}_4
   \right],
\label{eq4-26c}
\end{eqnarray}
where $S^{\alpha\beta}_m$ is given by equation (\ref{eq4-13}) without
any modification. In Table~\ref{tab4}, we show the values of skewness
parameters of the velocity field for the CDM-like models for
several values of $\Gamma R$.
\begin{table}
\begin{center}
\caption{3D skewness parameters of Gaussian-smoothed velocity
field for CDM-like models. The factor $-\logd$ is multiplied and the
values are almost independent on cosmological parameters.
\label{tab4}}
\begin{tabular}{c|cccccccc} \tableline\tableline
\hfil
$\Gamma R$ & $1.0$ & $2.0$ & $4.0$ & $8.0$ &
$16.0$ & $32.0$\\
\tableline
$-\logd S^{(0)}$ & $2.456$ & $2.232$ & $1.995$ & $1.773$ &
$1.581$ & $1.420$\\
$-\logd S^{(1)}$ & $2.551$ & $2.319$ & $2.065$ & $1.826$ &
$1.620$ & $1.448$\\
$-\logd S^{(2)}$ & $1.963$ & $1.877$ & $1.779$ & $1.683$ &
$1.601$ & $1.530$\\
\tableline
\end{tabular}
\end{center}
\end{table}
In this table, the value of $E_v$ is set as $-1/7$. 

For the power spectrum of the power-law form, 
\begin{eqnarray}
&&
   S^{(0)}(\nspec) =
   S^{(1)}(\nspec)
   =
   -\frac{1}{\logd}
   \left[
      3 F\left(\frac{\nspec+3}{2},\frac{\nspec+3}{2},\frac32;\frac14 \right)
      - (\nspec + 2 - 2E_v) 
      F\left(\frac{\nspec+3}{2},\frac{\nspec+3}{2}, \frac52;\frac14 \right)
   \right],
\label{eq4-27a}\\
&& 
   S^{(2)}(\nspec) =
   -\frac{3}{\logd}
   \left[
      F\left(\frac{\nspec+5}{2},\frac{\nspec+5}{2},\frac52;\frac14\right)
      - \frac15 (\nspec + 4 - 4E_v)
      F\left(\frac{\nspec+5}{2},\frac{\nspec+5}{2},\frac72;\frac14\right)
   \right].
\label{eq4-28c}
\end{eqnarray}
Several numerical values are shown in Table~\ref{tab5}, where $E_v =
-1/7$ is assumed.
\begin{table}
\begin{center}
\caption{3D skewness parameters of Gaussian-smoothed velocity
field for power-law models, which are the functions of the spectral
index $n$.
\label{tab5}}
\begin{tabular}{c|ccccccccc} \tableline\tableline
\hfil
$n$ & $-3.0$ & $-2.5$ & $-2.0$ & $-1.5$ & $-1.0$ & $-0.5$ & $0.0$ &
$0.5$ & $1.0$\\
\tableline
$-\logd S^{(0,1)}$ & $3.714$ & $3.250$ & $2.848$ & $2.498$ & $2.191$ & $1.918$ &
$1.670$ & $1.441$ & $1.222$\\
$-\logd S^{(2)}$ & $2.333$ & $2.170$ & $2.026$ & $1.900$ & $1.788$ &
$1.687$ & $1.595$ & $1.509$ & $1.425$\\
\tableline
\end{tabular}
\end{center}
\end{table}

\subsection{2D Projected Density Field}
\label{sec4-4}

The projection of the density field $\rho_{\rm p}$ defines the 2D
cosmic fields on the sky ($d=2$). Here, we derive the skewness
parameters for this field. The 2D projected density field with top-hat
kernel is investigated by \citet{ber95c}. Here, we are interested in
Gaussian kernel for our purpose.

In a Friedman-Lema\^{\i}tre universe, the comoving angular diameter
distance at a comoving distance $\chi$ is given by
\begin{eqnarray}
   S_K(\chi) = 
   \left\{
   \begin{array}{ll}
      \displaystyle
      \frac{\sinh\left(\chi\sqrt{-K}\right)}{\sqrt{-K}}, & K < 0, \\
      \chi, & K = 0, \\
      \displaystyle
      \frac{\sin\left(\chi\sqrt{K}\right)}{\sqrt{K}}, & K > 0,
   \end{array}
   \right.
\label{eq4-29}
\end{eqnarray}
depending on the sign of the spatial curvature $K = H_0^{\,2}(\Omega_0
+ \lambda_0 - 1)$. Thus, in spherical coordinates, projected
density field in 2D is given by
\begin{eqnarray}
   \rho_{\rm p}(\theta,\phi) = 
   \int d\chi {S_K}^2 (\chi) n(\chi)
   \rho(\chi,\theta,\phi;\tau_0-\chi),
\label{eq4-30}
\end{eqnarray}
where $n(\chi)$ is the selection function without volume factor,
normalized as $\int d\chi S_K^{\,2}(\chi) n(\chi) = 1$, and
$\rho(\chi,\theta,\phi; \tau)$ is the 3D comoving density
field\footnote{The comoving density field is defined by the density
per unit comoving volume and thus satisfy $\bar{\rho} =$ constant.}.
The present value of the conformal time $\tau=\int dt/a$ is $\tau_0$.

The projected density contrast is defined by $\rho_{\rm
p}/\bar{\rho}_{\rm p} - 1$, where one can see $\bar{\rho}_{\rm p} =
\bar{\rho}$. We identify the projected density contrast with the 2D
($d=2$) field $f$. Since the smoothing angle $\thf$ is much smaller
than $\pi$ in most of the interested cases, we consider the small
patch of the sky of the vicinity of the polar axis, $\theta \ll 1$.
With this approximation, we introduce the variables $\theta_1 =
\theta\cos\phi$, and $\theta_2 = \theta\sin\phi$, which are considered
as 2D Euclidean coordinates, $\bfth$. Therefore, the projection
equation is given by
\begin{eqnarray}
   f(\bfth) = 
   \int d\chi S_K^{\,2}(\chi) n(\chi)
   \delta_{\rm 3D} (\chi, S_K(\chi)\bfth;\tau_0-\chi),
\label{eq4-31}
\end{eqnarray}
where $\delta_{\rm 3D}(\bfx, \tau) = \rho/\bar{\rho} - 1$ is the
density contrast at comoving coordinates $\bfx$ and conformal time
$\tau$.

The power spectrum and the bispectrum for the above projected field
are given by the Limber's equations (\ref{eqa2-2}) and (\ref{eqa2-7})
in Appendix~\ref{app2}, with $q(\chi) = n(\chi)$:
\begin{eqnarray}
&&
   P_{\rm 2D}(\omega) =
   \int d\chi S_K^{\,2}(\chi) n^2(\chi)
   P_{\rm 3D}(\frac{\omega}{S_K(\chi)};\tau_0-\chi),
\label{eq4-33a}\\
&&
   B_{\rm 2D}(\omega_1,\omega_2,\omega_3) =
   \int d\chi S_K^{\,2}(\chi) n^3(\chi)
   B_{\rm 3D}(\frac{\omega_1}{S_K(\chi)},\frac{\omega_2}{S_K(\chi)},
      \frac{\omega_2}{S_K(\chi)};\tau_0-\chi),
\label{eq4-33b}
\end{eqnarray}
where $P$, $B$ are 2D projected power spectrum and bispectrum,
respectively, of the field $f$, and $P_{\rm 3D}$, $B_{\rm 3D}$ are 3D
power spectrum and bispectrum, respectively. The 3D power spectrum and
the 3D bispectrum are evaluated by the second order perturbation
theory. They are similar to equations (\ref{eq4-7a}) and
(\ref{eq4-7b}), but we have to take into account the time-dependence
here. They are given by
\begin{eqnarray}
&&
   P_{\rm 3D}(k;\tau_0 - \chi) = D^2(\chi) \PL(k)
\label{eq4-34a}\\
&&
   B_{\rm 3D}(k_1, k_2, k_3;\tau_0 - \chi)
   =
   D^4(\chi) 
   \left[
      1 + E(\chi) + 
      \left(\frac{k_2}{k_1} + \frac{k_1}{k_2}\right)
      \frac{\bfk_1\cdot\bfk_2}{k_1 k_2} +
      (1 - E(\chi))
      \left(\frac{\bfk_1\cdot\bfk_2}{k_1 k_2}\right)^2
   \right] \PL(k_1) \PL(k_2)
\nonumber\\
&&\qquad\qquad\qquad\qquad\qquad\quad
   +\, {\rm cyc.(1,2,3)},
\label{eq4-34b}
\end{eqnarray}
where $D(\chi)$ is the linear growth factor at conformal lookback time
$\chi$, (i.e., at conformal time $\tau = \tau_0 - \chi$), which is
normalized as $D(0) = 1$. The following fitting formula \citep{car92}
is useful:
\begin{eqnarray}
   D \approx
   \frac{a\Omega}{\Omega_0}\cdot
   \frac{
      {\Omega_0}^{4/7} - \lambda_0
      + (1 + \Omega_0/2)(1 + \lambda_0/70)
   }{
      \Omega^{4/7} - \lambda
      + (1 + \Omega/2)(1 + \lambda/70)
   },
\label{eq4-34-1}
\end{eqnarray}
where $\Omega$ and $\lambda$ are time-dependent cosmological
parameters at conformal lookback time $\chi$. The variable $E(\chi)$
is a weak function of time and cosmology, and for Einstein-de Sitter
universe, $E = 3/7$. This quantity $E$ is the same we used in 3D
density field, but here we also take into account the time-dependence.
It is accurately approximated by
\begin{eqnarray}
   E \approx \frac37 \Omega^{-1/30} -
   \frac{\lambda}{80}
   \left(1 - \frac32 \lambda \log_{10}\Omega\right).
\label{eq4-35}
\end{eqnarray}

The variance parameters of the smoohted projected field are
given by
\begin{eqnarray}
   {\sigma_j}^2(\thf) =
   \int\frac{\omega d\omega}{2\pi}
   \omega^{2j} P(\omega) W^2(\omega\thf) =
   \frac{1}{\thf^{2j+2}}
   \int d\chi S_K^{\,2}(\chi) n^2(\chi) D^2(\chi)
   {\Sigma_j}^2\left[S_K(\chi)\thf\right],
\label{eq4-36}
\end{eqnarray}
where
\begin{eqnarray}
   {\Sigma_j}^2(R) =
   R^{2j+2} \int \frac{kdk}{2\pi} k^{2j} 
   \PL(k) W^2(kR) =
   \int \frac{ldl}{2\pi} l^{2j} 
   \PL\left(\frac{l}{R}\right) W^2(l).
\label{eq4-37}
\end{eqnarray}
The skewness parameters of the smoothed projected field are
given by
\begin{eqnarray}
&&
   S^{(a)}(\thf) =
   \frac{1}{{\sigma_0}^4 \thf^4}
   \left(\frac{\sigma_0}{\sigma_1\thf}\right)^{2a}
   \int d\chi {S_K}^2(\chi) n^3(\chi) D^4(\chi)
   {\Sigma_0}^{4-2a}\left[S_K(\chi)\thf\right]
   {\Sigma_1}^{2a}\left[S_K(\chi)\thf\right]
   C^{(a)}\left[S_K(\chi)\thf\right],
\label{eq4-38}
\end{eqnarray}
where
\begin{eqnarray}
&&   C^{(a)}(R) =
   \frac{1}{{\Sigma_0}^4}
   \left(\frac{\Sigma_0}{\Sigma_1}\right)^{2a}
   \int \frac{l_1dl_1}{2\pi} \frac{l_2dl_2}{2\pi}
   \PL \left(\frac{l_1}{R}\right)
   \PL \left(\frac{l_2}{R}\right)
   W^2(l_1) W^2(l_2)
   \widetilde{C}^{(a)}(l_1, l_2),
\label{eq4-39}
\end{eqnarray}
and
\begin{eqnarray}
&&
   \widetilde{C}^{(a)}(l_1,l_2) =
   \frac{3}{2\pi}
   \int_{-1}^1 \frac{d\mu}{\sqrt{1 - \mu^2}}
   \left[
      1+E + 
      \left(\frac{l_2}{l_1} + \frac{l_1}{l_2}\right) \mu +
      (1-E) \mu^2
   \right]
   \frac{W\left(\sqrt{l_1^{\,2} + l_2^{\,2} + 2 l_1 l_2 \mu} \right)}
      {W(l_1) W(l_2)}
\nonumber\\
&&\qquad\qquad\quad \times\,
   \left\{
   \begin{array}{ll}
      \displaystyle
      2, & a=0,\\
      {l_1}^2 + {l_2}^2 + l_1 l_2 \mu, & a=1,\\
      4\, {l_1}^2 {l_2}^2 \left( 1 - \mu^2 \right), & a=2,
   \end{array}
   \right.
\label{eq4-40}
\end{eqnarray}
So far the smoothing function is arbitrary. For a general smoothing
function, the above equations can be numerically integrated to obtain
the skewness parameters for each model of the power spectrum. In the
following, we adopt the Gaussian smoothing function, $W(l) =
\exp(-l^2/2)$. For this smoothing function, the equation
(\ref{eq4-10}) holds even for this 2D case. In this case, the
following integral representation of the modified Bessel function
$I_\nu(z)$ for $\nu = 0$ is useful:
\begin{eqnarray}
   I_0(z) = 
   \frac{1}{\pi}
   \int_{-1}^1 \frac{d\mu}{\sqrt{1 - \mu^2}} e^{-z\mu}.
\label{eq4-42}
\end{eqnarray}
Actually, the derivatives of the above equation, 
\begin{eqnarray}
   \frac{1}{\pi}
   \int_{-1}^1 \frac{d\mu}{\sqrt{1 - \mu^2}} \mu^m e^{-z\mu} =
   \left(-\frac{d}{dz}\right)^m I_0(z),
\label{eq4-43}
\end{eqnarray}
are sufficient to perform the angular integration in equation
(\ref{eq4-40}). Moreover, one can use the property
of the Bessel function, $I_0' = I_1$, and $I_m' = (I_{m-1} +
I_{m+1})/2$ to obtain formulas,
\begin{eqnarray}
&&
   \frac{1}{\pi}
   \int_{-1}^1 \frac{d\mu}{\sqrt{1 - \mu^2}} \mu e^{-z\mu} =
   - I_1(z),
\label{eqd42-44a}\\
&&
   \frac{1}{\pi}
   \int_{-1}^1 \frac{d\mu}{\sqrt{1 - \mu^2}} \mu^2 e^{-z\mu} =
   \frac12 I_0(z) + \frac12 I_2(z),
\label{eq4-44b}\\
&&
   \frac{1}{\pi}
   \int_{-1}^1 \frac{d\mu}{\sqrt{1 - \mu^2}} \mu^3 e^{-z\mu} =
   - \frac34 I_1(z) - \frac14 I_3(z),
\label{eq4-44c}\\
&&
   \frac{1}{\pi}
   \int_{-1}^1 \frac{d\mu}{\sqrt{1 - \mu^2}} \mu^4 e^{-z\mu} =
   \frac38 I_0(z) + \frac12 I_2(z) + \frac18 I_4(z).
\label{eq4-44d}
\end{eqnarray}
{}From these formulas, equation (\ref{eq4-40})
reduce to 
\begin{eqnarray}
&&
   \widetilde{C}^{(0)} = 
   \frac32
   \left[
      (3+E) I_0(l_1 l_2) -
      2 \left(\frac{l_2}{l_1} + \frac{l_1}{l_2}\right) I_1(l_1 l_2) +
      (1-E) I_2(l_1 l_2)
   \right],
\label{eq4-45a}\\
&&
   \widetilde{C}^{(1)} = 
   \frac34
   \left\{
      (4+E) (l_1^{\,2} + l_2^{\,2}) I_0(l_1 l_2) -
      \left[
         \frac{15 + E}{2} l_1 l_2 + 
         2
         \left(
            \frac{l_1^{\,3}}{l_2} + \frac{l_2^{\,3}}{l_1}
         \right)
      \right] I_1(l_1 l_2)
   \right.
\nonumber\\
&&\qquad\qquad\qquad
   \left. +\;
      (2-E) (l_1^{\,2} + l_2^{\,2}) I_2(l_1 l_2) -
      \frac{1 - E}{2} l_1 l_2 I_3(l_1 l_2)
   \right\},
\label{eq4-45b}\\
&&
   \widetilde{C}^{(2)} = 
   \frac34
   \left[
      (5+3E) l_1^{\,2} l_2^{\,2} I_0(l_1 l_2) -
      2 l_1 l_2(l_1^{\,2} + l_2^{\,2}) I_1(l_1 l_2) -
      4(1+E) l_1^{\,2} l_2^{\,2} I_2(l_1 l_2)
   \right.
\nonumber\\
&&\qquad\qquad
   \left. +\;
      2 l_1 l_2 (l_1^{\,2} + l_2^{\,2}) I_3(l_1 l_2) -
      (1-E) l_1^{\,2} l_2^{\,2} I_4(l_1 l_2)
   \right].
\label{eq4-45c}
\end{eqnarray}
At this point, defining
\begin{eqnarray}
&&
   C^{\alpha\beta}_m(R) =
   \frac{1}{{\Sigma_0}^{4}}
   \left(\frac{\Sigma_0}{\Sigma_1}\right)^{\alpha+\beta-2}
   \int \frac{l_1dl_1}{2\pi} \frac{l_2dl_2}{2\pi}
   \PL\left(\frac{l_1}{R}\right)
   \PL\left(\frac{l_2}{R}\right)
   e^{-l_1^{\,2}-l_2^{\,2}} {l_1}^{\alpha-1} {l_2}^{\beta-1}
   I_m(l_1 l_2),
\label{eq4-46}
\end{eqnarray}
the equation (\ref{eq4-39}) reduces to
\begin{eqnarray}
&&
   C^{(0)} = 
   \frac32
   \left[
      (3+E) C^{11}_0
      - 4 C^{02}_1
      + (1-E) C^{11}_2
   \right],
\label{eq4-47a}\\
&&
   C^{(1)} = 
   \frac34
   \left[
      2(4+E) C^{13}_0
      - \frac{15 + E}{2} C^{22}_1
      - 4 C^{04}_1
      + 2(2-E) C^{13}_2
      - \frac{1 - E}{2} C^{22}_3
   \right],
\label{eq4-47b}\\
&&
   C^{(2)} =
   \frac34
   \left[ 
      (5+3E) C^{33}_0
      - 4 C^{24}_1
      - 4(1 + E) C^{33}_2
      + 4C^{24}_3
      - (1 - E) C^{33}_4
   \right].
\label{eq4-47c}
\end{eqnarray}
The two-dimensional integration of equation (\ref{eq4-46}) is
performed only once as a function of $R$. The result is stored as a
table, and is used in one-dimensional numerical integration of
equation (\ref{eq4-38}) for finally obtaining the skewness parameters
in 2D projected density fields. Functions $C^{(a)}$ for CDM-like
models are given in Table~\ref{tab6} which are functions of $\Gamma
R$.
\begin{table}
\begin{center}
\begin{tabular}{c|cccccccccccc} \tableline\tableline
\hfil
$\Gamma R$ & $0.25$ & $0.5$ & $1.0$ & $2.0$ & $4.0$ & $8.0$ &
$16.0$ & $32.0$ & $64.0$\\
\tableline
$C^{(0)}$ & $4.544$ & $4.311$ & $4.020$ & $3.685$ & $3.345$ & $3.044$ &
$2.797$ & $2.613$ & $2.504$\\
$C^{(1)}$ & $4.885$ & $4.614$ & $4.279$ & $3.890$ & $3.492$ & $3.143$ &
$2.863$ & $2.653$ & $2.523$\\
$C^{(2)}$ & $3.723$ & $3.660$ & $3.589$ & $3.518$ & $3.467$ & $3.453$ &
$3.479$ & $3.532$ & $3.584$\\
\tableline
\end{tabular}
\end{center}
\caption{Functions $C^{(a)}(R)$ for CDM-like models, which are the
functions of the product of shape parameter $\Gamma$ and smoothing
length $R$. The parameter $E$ is set as $E=3/7$.
\label{tab6}
}
\end{table}

When the 3D power spectrum is given by a power-law form of equation
(\ref{eq4-16}), the integration by wave length $l_1$, $l_2$ can be
analytically performed as in the 3D case. In fact, the parameter
$\Sigma_j$ of equation (\ref{eq4-37}) with Gaussian smoothing $W(l) =
e^{-l^2/2}$ is given by
\begin{eqnarray}
   {\Sigma_j}^2(R) = 
   \frac{A}{4\pi R^\nspec}
   \Gamma\left(\frac{\nspec+2j+2}{2}\right),
\label{eq4-48}
\end{eqnarray}
and the variance parameters are given by
\begin{eqnarray}
   {\sigma_j}^2(\thf) =
   \frac{A}{4\pi \thf^{\nspec+2j+2}}
   \Gamma\left(\frac{\nspec+2j+2}{2}\right)
   \int d\chi {S_K}^{2-\nspec}(\chi) n^2(\chi) D^2(\chi).
\label{eq4-48-2}
\end{eqnarray}
With similar technique used in 3D density field, 
\begin{eqnarray}
&&
   C^{\alpha\beta}_m =
   \frac{1}{2^m m!}
   \left(\frac{n+2}{2}\right)^{1-(\alpha+\beta)/2}
   \left(\frac{n+2}{2}\right)_{(\alpha+m-1)/2}
   \left(\frac{n+2}{2}\right)_{(\beta+m-1)/2}
\nonumber\\
&&\qquad\qquad\qquad\qquad \times
   F\left(
      \frac{n+\alpha+m+1}{2},\frac{n+\beta+m+1}{2},m+1;\frac14
   \right).
\label{eq4-49}
\end{eqnarray}
The equations (\ref{eq4-47a})--(\ref{eq4-47c}) and the above equation
finally give the values of $C^{(a)}$. The recursion relations of
equations (\ref{eq4-21a}) and (\ref{eq4-21b}) simplify the result:
\begin{eqnarray}
&&
   C^{(0)}(\nspec) =
   C^{(1)}(\nspec)
   =
   3 F\left(\frac{\nspec+2}{2},\frac{\nspec+2}{2},1;\frac14\right) -
   \frac32 (\nspec+1-E)
   F\left(\frac{\nspec+2}{2},\frac{\nspec+2}{2},2;\frac14\right),
\label{eq4-50a}\\
&&
   C^{(2)}(\nspec) =
   3 F\left(\frac{\nspec+4}{2},\frac{\nspec+4}{2},2;\frac14\right) -
   \frac{3(\nspec+3-3E)}{4}
   F\left(\frac{\nspec+4}{2},\frac{\nspec+4}{2},3;\frac14\right).
\label{eq4-50c}
\end{eqnarray}
These functions are independent on scale $R$ in the power-law case,
but dependent on power-law index, $\nspec$. Again, $C^{(0)}(\nspec)$
and $C^{(1)}(\nspec)$ are identical only in the power-law case.
Numerical values are given in Table~\ref{tab7}, where $E=3/7$ is
assumed.
\begin{table}
\begin{center}
\begin{tabular}{c|ccccccccc} \tableline\tableline
\hfil
$\nspec$ & $-3.0$ & $-2.5$ & $-2.0$ & $-1.5$ & $-1.0$ & $-0.5$ & $0.0$ &
$0.5$ & $1.0$\\
\tableline
$C^{(0,1)}$ & $6.949$ & $5.965$ & $5.143$ & $4.457$ & $3.885$ & $3.409$ &
$3.014$ & $2.688$ & $2.421$\\
$C^{(2)}$ & $4.090$ & $3.863$ & $3.687$ & $3.560$ & $3.478$ & $3.443$ &
$3.445$ & $3.518$ & $3.635$\\
\tableline
\end{tabular}
\end{center}
\caption{Functions $C^{(a)}$ for power-law models, which are the
functions of the spectral index $n$. The parameter $E$ is set as
$E=3/7$. 
\label{tab7}
}
\end{table}
For the power-law case, the skewness parameters of equation
(\ref{eq4-38}) reduces to the following simple form:
\begin{eqnarray}
   S^{(a)}(\nspec) =
   \frac{\displaystyle
       \int d\chi \left[S_K(\chi)\right]^{2-2\nspec} n^3(\chi) D^4(\chi)}
      {\displaystyle
       \left\{
         \int d\chi \left[S_K(\chi)\right]^{2-\nspec} n^2(\chi) D^2(\chi)
      \right\}^2}\,
    C^{(a)}(\nspec).
\label{4-50-2}
\end{eqnarray}

It is interesting to compare the results with those of top-hat
smoothing. According to \citet{ber95c}, the top-hat smoothing gives
$C^{(0)}(\nspec) = 36/7 + 3(\nspec + 2)/2$ for power-law case.
Comparing this expression with our Table~\ref{tab7}, they roughly
agree with each other, and have similar behavior with spectral index.
In detail, the Gaussian smoothing gives more or less larger values.
For example, top-hat smoothing gives $C^{(0)} = 5.14, 3.64, 2.14$ for
$\nspec=-2, -1,0$, respectively, while Gaussian smoothing gives
$C^{(0)} = 5.14, 3.89, 3.01$, respectively. It is reasonable that the
$C^{(0)}$ of top-hat smoothing differs from that of Gaussian smoothing
for large spectral index, because the top-hat smoothing gathers more
power on small scales than Gaussian smoothing.

\subsection{Weak Lensing Field}
\label{sec4-5}

The local convergence field of the weak lensing is commonly used for
studying the large-scale structure of the universe
\citep[e.g.,][]{kai98,bar99}. Assuming the situation where Limber's
equation (see Appendix~\ref{app2}) and also the Born approximation
\citep{kai98} hold, the following correspondence between the
convergence field $\kappa$ and the 3D density contrast $\delta_{\rm
3D}$ is useful \cite[e.g.,][]{mel99}:
\begin{eqnarray}
   \kappa(\bfth) = \frac32 {H_0}^2 \Omega_0
   \int_0^\infty d\chi' {S_K}^2(\chi') n(\chi')
   \int_0^{\chi'} d\chi
   \frac{S_K(\chi) S_K(\chi' - \chi)}{a(\chi) S_K(\chi')}
   \delta_{\rm 3D}
   \left[\chi',\bfth S_K(\chi');\tau_0 - \chi'\right],
\label{eq4-60}
\end{eqnarray}
where $a(\chi)$ is the scale factor at conformal lookback time $\chi =
\tau_0 - \tau$. The above equation is reduced to exactly the same form
as the projected field of equation (\ref{eq4-31}), but with the
substitution,
\begin{eqnarray}
   n(\chi) \rightarrow
   n_{\rm wl}(\chi) = 
   \frac32 \frac{{H_0}^2 \Omega_0}{a(\chi)}
   \int_\chi^\infty d\chi' {S_K}^2(\chi')
   \frac{S_K(\chi'-\chi)}{S_K(\chi')S_K(\chi)}
   n(\chi').
\label{eq4-62}
\end{eqnarray}
However, one should note that this substitution is valid only under
the assumption of the Born approximation. Although the effect of the
Born approximation on the skewness is known to be weak \citep{ber97},
the validity of the Born approximation in general situation has not
been tested in detail \citep{mel99}. There is some subtlety which
could arise when various combinations of skewness are considered.

\subsection{The Biases}
\label{sec4-6}

The above expressions of the skewness parameters are for unbiased
fields. The skewness parameters for biased fields are non-trivial.
They depend on the details of the biasing scheme in the real universe
which is poorly known so far. However, the skewness parameters are
well-defined quantities, so that they are calculated from the first
principle once the biasing scheme is given.

Perhaps, one of the simplest, yet non-trivial case is the local,
deterministic biasing. In this case, we can follow \citet{fry93} to
obtain the perturbative expansion of the biasing:
\begin{eqnarray}
   \delta_{\rm g} =
   b\, \delta + 
   \frac{b_2}{2}
   \left(\delta^2 - \left\langle\delta^2\right\rangle\right) +
   \cdots,
\label{eq4-70}
\end{eqnarray}
where $\delta_{\rm g}$ and $\delta$ are the galaxy and mass density
contrast, respectively. The spatial dimension is arbitrary, so that
$\delta_{\rm g}$, $\delta$ can be functions in either 1D, 2D, or 3D
space. The biased variance in lowest order is given by
\begin{eqnarray}
   \sigma_{\rm 0,g} = b \sigma_0.
\label{eq4-70-1}
\end{eqnarray}
It is also straightforward to calculate skewness parameters. After
some algebra, all skewness parameters are shown to transform in a same
way:
\begin{eqnarray}
   S^{(a)}_{\rm g} = \frac{S^{(a)}}{b} + \frac{3 b_2}{b^2},
\label{eq4-71}
\end{eqnarray}
irrespective of spatial dimensions, $d$, and of kinds of skewness
parameters, $a=0,1,2$. In this framework, the parameters $b$ and $b_2$
are needed. A possible way to determine these parameters is to measure
the variance ${\sigma_{\rm g}}^2$ and the skewness $S_{\rm g}$ from
the observation. Theoretical models predict the variance ${\sigma_{\rm
m}}^2$ and the skewness $S_{\rm m}$ of the mass distribution. We
obtain biasing parameters by $b = \sigma_{\rm g}/\sigma_{\rm m}$ and
$b_2 = b(b S_{\rm g} - S_{\rm m})/3$. The derivatives of the skewness,
$S^{(1)}_{\rm g}$, $S^{(2)}_{\rm g}$, are then obtained from equation
(\ref{eq4-71}).

In case of level-crossing, and genus statistics as functions of scaled
threshold $\tnu$, the second-order corrections in equation
(\ref{eq2-207a})--(\ref{eq2-207c}) only depends on the difference of
the skewness parameters times the variace. Quite remarkably, this type
of second-order correction term does not depend on bias parameter at
all:
\begin{eqnarray}
   \left[S^{(a)}_{\rm g} - S^{(0)}_{\rm g}\right] \sigma_{\rm 0,g} =
   \left[S^{(a)} - S^{(0)}\right] \sigma_{\rm 0}
\label{eq4-72}
\end{eqnarray}
Thus, the second-order nonlinear corrections for level-crossing and
genus statistics of locally biased field are exactly the same as
that of unbiased mass density field. This can be considered as an
advantage of the scaled threshold $\tnu$ in these statistics, 
because the biasing is one of the most embarrasing uncertainty in the
analysis of galaxy distribution. However, one should note this
result is derived under the local biasing scheme. So far the locality
of the bias is not guaranteed in general. 

Stochastic argument \citep{dek99} of the biasing is more complicated
for derivative skewness $S^{(1)}$ and $S^{(2)}$ than for usual
skewness $S^{(0)}$, because they involve the correlation between field
derivatives. The phenomenological nature of the stochastic biasing
requires many parameters which is not calculable from first
principles. Therefore, the stochastic biasing sheme does not
effectively work for our problem. More physical treatment of the
biasing schemes of galaxy formation is needed, in which case the
nonlocality of the bias could also be important \citep{mat99}.

\setcounter{equation}{0}
\section{IMPLICATIONS OF SECOND-ORDER RESULTS}
\label{sec5}

The perturbative calculations offer valuable aspects of weakly
non-linear evolution of the various statistics without laborious
parameter survey by numerical simulations. In this paper, we obtain
the lowest nonlinear corrections to relatively popular statistics of
smoothed cosmic fields. Using these results, it is interesting to see
how the weakly nonlinear effect tends to distort the Gaussian
prediction for those exemplified statistics.

\begin{figure}
\epsscale{0.65} \plotone{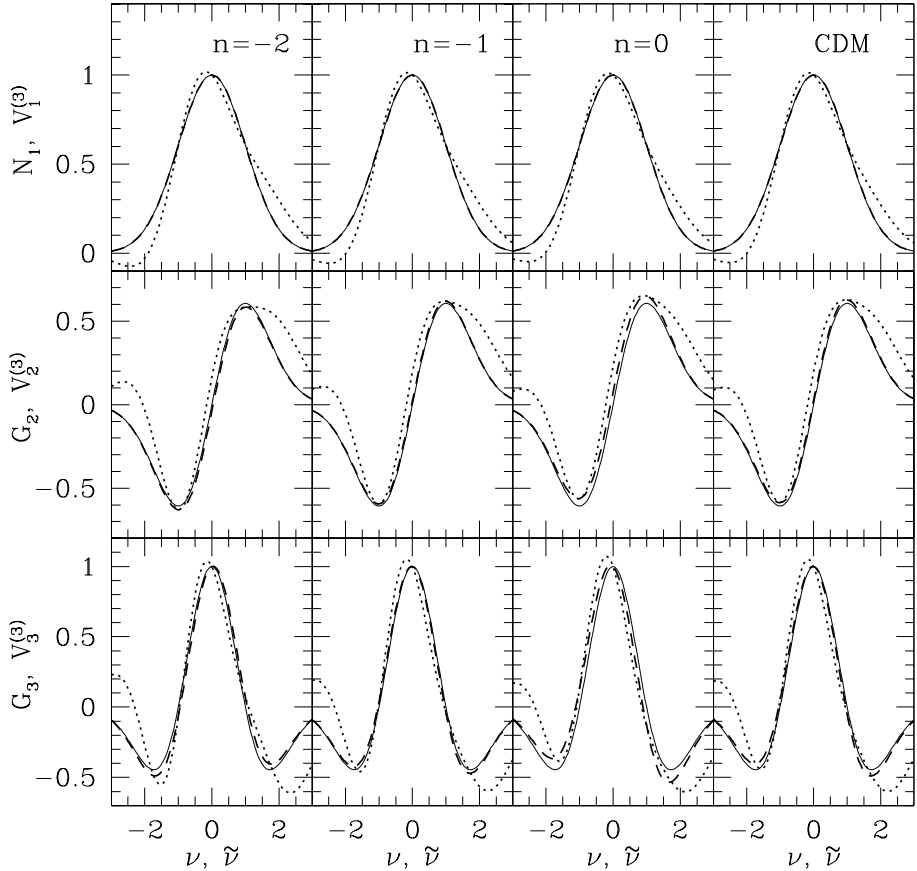} \figcaption[f1.eps]{The 3D genus,
$G_3$, the 2D genus $G_2$, the level-crossing statistic $N_1$ and the
Minkowski functionals $V^{(3)}_k$ of the 3D density field. All curves
are appropriately normalized. The variance is set as $\sigma_0 = 0.3$.
Solid lines: Gaussian predictions, dotted lines: second order
predictions in terms of density threshold, $\nu$, dashed lines: second
order predictions in terms of volume-fraction threshold, $\tnu$. The
initial density fluctuation spectrum is given by the power-law with
$n=-2,-1,0$ and also by the CDM-like model with smoothing length
$R=4/\Gamma$ (see text).
\label{fig1}}
\end{figure}

In Figure~\ref{fig1}, various statistics for 3D density field is
shown.
The amplitude of each statistic is appropriately normalized as we are
interested in the deviation from the Gaussian prediction. If we
neglect the normalization, Minkowski functionals of $k=1,2,3$ are
equivalent to the statistics $N_1$, $G_2$, and $G_3$, respectively, so
that they degenerate in this figure (we should note the sign of
$V^{(3)}_3$ is inverted).

The rms $\sigma_0$, which is considered as a weakly nonlinear
parameter, is set $\sigma_0=0.3$. A limit $\sigma_0 \rightarrow 0$
corresponds to the prediction of the linear theory, which is given by
thin solid lines in the figure. This linear prediction is equivalent
to the Gaussian fluctuations, because we assume the initial density
field is random Gaussian. 

In general, the curves of statistics plotted against the direct
density threshold, $\nu$ (dotted lines), exhibit considerable
deviations from Gaussian predictions. The overall tendency does not
depend much on the shape of the spectrum we consider here, i.e.,
power-law spectrum with index $-2$, $-1$, $0$, and CDM model with
smoothing length $R = 4/\Gamma$, where $\Gamma$ is the shape parameter
of the CDM spectrum. They are consistent with the so-called meat-ball
shift, which means that there are more isolated regions in a nonlinear
field than in a Gaussian field for a fixed threshold. In fact, $N_1$,
$G_2$, $-G_3$ of high value of threshold, e.g., $\nu\sim 2$, virtually
correspond to the number of isolated regions, and each figure shows
that the number is indeed increased by weakly nonlinear evolution.

The weakly nonlinear formula for the genus curve against the density
threshold, $G_3(\nu)$, which was first derived by \citet{mat94} has
been compared with numerical simulations in literatures.
\citet{mat96b} shows the good agreement of the analytic prediction
with the simulations for various spectra. \citet{col00} compared the
prediction of the genus curve against direct $\nu$ with the simulated
SDSS data. Unfortunately, in their published paper, they have
transcribing errors, which incorrectly made the perturbation theory
considerably disagree with their data. There are also ambiguity in
their comparison on biasing which alter the values of skewness
parameters. One can guess the biasing effect on skewness parameters by
equation (\ref{eq4-70}). They chose the peak particles as galaxies,
and the linear biasing parameter is inferred as $b \sim 1.3$, but the
nonlinear parameter $b_2$ is not obvious in their work. Some
literatures indicate the skewness $S^{(0)}$ of peaks are roughly given
by 1--2 \citep{wat94,pli95} but for highly biased peaks $b\sim 2$. If
we adopt $b_2 = -0.5$, the skewness is given by $S^{(0)}=1.8$ which is
not unreasonable. If it is the case for their simulation, the
perturbative prediction and their data completely agree with each
other. The $\chi^2$-value per degrees of freedom reduces to only 1.03
(private communication with W. N. Colley \& D. H. Weinberg).
Obviously, we have to further investigate the biasing effects in
numerical simulations to obtain a conclusive result.

Most of the topological analyses of the previous work use the scaled
threshold $\tnu$. The dashed lines in Fig.~\ref{fig1} shows the
corresponding curves. The deviations from the linear theory is
dramatically reduced. This fact is empirically known by the analyses
of numerical simulations \citep{got86,got87}. The reason for this
reduction is mathematically due to the closeness of the values of
skewness parameters $S^{(a)}$, ($a=0,1,2$), since all the terms of the
nonlinear corrections in equations (\ref{eq2-207a})--(\ref{eq2-207c})
depend only on $S^{(a)} - S^{(0)}$ ($a=1,2$). For the hierarchical
model of equation (\ref{eq4-1}), they are exactly zero, that means
there is not any (second order) nonlinear correction for the
hierarchical model. Since the hierarchical model is known to roughly
approximate the nonlinear evolution, it is not surprising that more
realistic fields have only small corrections of nonlinearity if they
are plotted against the volume-fraction threshold, $\tnu$.

For power-law models, $S^{(0)}$ and $S^{(1)}$ are exactly the same.
That makes the nonlinear correction for $N_1$ or $V^{(3)}_1$ exactly
vanishes. Thus, the nonlinear corrections for other statistics are
arisen by the difference between $S^{(0)}$ and $S^{(2)}$. For the CDM
model, there still is a difference between $S^{(0)}$ and $S^{(1)}$,
but it is relatively small as seen from the Table~\ref{tab2}.

As for the topological statistics, $G_2$, $G_3$, $V^{(3)}_2$ and
$V^{(3)}_3$, deviations of curves against $\tnu$ from the linear
theory prediction depend on the underlying spectrum through
differences of skewness parameters. The weakly nonlinear effect for
redder spectrum of $\nspec=-2$ induces a sponge-like shift, that means
the number of holes in isolated regions increases. On the other hand,
the bluer spectrum of $\nspec=0$ indicates a meat-ball shift. These
tendencies are qualitatively in agreement with the numerical results
\cite[e.g.][]{ryd89,mel89,par91}.

It is not trivial to estimate errors expected in observationally
estimating statistics of smoothed cosmic fields. The observational
errors mainly consists of the shot noise and the cosmic variace. The
systematic comparison with simulations should be used for the error
estimates. Unfortunately, previous earlier simulations does not have
enough resolution to be quantitatively compared in weakly nonlinear
regime. \citet{can98} use much larger simulations than those earlier
work, and gives the genus curve in weakly nonlinear regime in their
Fig.~9. Relative depths of left and right troughs in the genus curve
show slightly meat-ball shift, which is consistent with our
prediction. Although their variation of smoothing length is rather
limited, their plots ensures the slight meat-ball shift predicted by
the perturbation theory can definitely be observed. \cite{col00} plot
the 2D genus curve, but they use the thin slice which suffers strongly
nonlinear effects so that the quantitative comparison is not
appropriate with weakly nonlinear results. They report the meat-ball
shift in 2D genus, which is the same direction of the weakly nonlinear
correction besides amplitude. We find there is stll neccesity of
proper, systematic comparison between the perturbation theory and
numerical simulations in weakly nonlinear regime.

The decisive factor between sponge-like shift and meat-ball shift is
the sign and amplitude of $S^{(2)} - S^{(0)}$, since $S^{(1)} -
S^{(0)}$ is exactly or approximately zero for wide range of models
like power-law, or CDM-like models. If the factor $S^{(2)} - S^{(0)}$
is positive, the meat-ball shift takes place. If this factor is
negative, the sponge-like shift occurs. For power-law case, that
factor is positive for $\nspec > -1.4$, and is negative for $\nspec <
-1.4$. The amplitude of this factor times the amplitude of the
nonlinearity parameter, $(S^{(2)} - S^{(0)})\sigma_0$ determines the
amplitude of the meat-ball shift in the genus curve.

Prominent shifts are seen around the two troughs of the genus curve at
$\tnu = \pm \sqrt{3}$. At these troughs, the factor $S^{(1)} -
S^{(0)}$ does not contribute to the genus curve since the accompanying
factor $H_3(\tnu) = \tnu^3 - \tnu$ vanishes. Therefore, one can almost
completely characterize the shifts by a factor $(S^{(2)} -
S^{(0)})\sigma_0$, which we call the 'genus asymmetry parameter',
around the troughs. We propose this genus asymmetry parameter as a
theory-motivated parameter of the genus asymmetry. This factor is
observationally determined by fitting the shape of the genus curve by
a form,
\begin{eqnarray}
   G(\tnu) \propto
   H_2(\tnu) + A H_3(\tnu) + B H_1(\tnu),
\label{eq5-0}
\end{eqnarray}
where $B$ is the genus asymmetry parameter. In this fitting, $A$ is
much smaller if the non-Gaussian features are from purely
gravitational evolution of the initial Gaussian field, and if the
power spectrum is smooth enough as CDM models. The genus asymmetry
parameter for CDM models with shape parameter $\Gamma$, normalized by
$\sigma_8$ is listed for various smoothing length in Table~\ref{tab8}.
\begin{deluxetable}{cccccccccc}
\tablecolumns{10}
\tablewidth{0pt}
\tablecaption{Genus asymmetry parameter $(S^{(2)} - S^{(0)})\sigma_0$
for a CDM model with shape parameter $\Gamma$. The power spectrum is
normalized by $\sigma_8 = 1$. The parameter $E$ is set as $E=3/7$.
\label{tab8}}
\tablehead{
\colhead{} &\colhead{} & \multicolumn{8}{c}{Smoothing length $R\ [\himpc]$}\\
\colhead{$\Gamma$} &\colhead{} & \colhead{$6.0$} & \colhead{$8.0$} &
\colhead{$10.0$} & \colhead{$12.0$} & \colhead{$14.0$} &
\colhead{$16.0$} & \colhead{$18.0$} & \colhead{$20.0$}} 
\startdata
$0.1$ & ... & $-0.101$ & $-0.045$ & $-0.011$ & $0.011$ & $0.025$ & $0.034$ &
$0.041$ & $0.045$ \\
$0.2$ & ... & $0.017$ & $0.054$ & $0.071$ & $0.078$ & $0.081$ & $0.080$ &
$0.079$ & $0.076$ \\
$0.3$ & ... & $0.088$ & $0.107$ & $0.111$ & $0.108$ & $0.102$ & $0.096$ &
$0.090$ & $0.084$ \\
$0.4$ & ... & $0.138$ & $0.141$ & $0.134$ & $0.123$ & $0.113$ & $0.102$ &
$0.093$ & $0.085$ \\
$0.5$ & ... & $0.175$ & $0.165$ & $0.149$ & $0.132$ & $0.118$ & $0.105$ &
$0.094$ & $0.085$ \\
\enddata
\end{deluxetable}
Since the factor $S^{(2)} - S^{(0)}$ increase with the spectral index,
as seen from Table~\ref{tab3}, this factor also increase with the
smoothing length for CDM model on scales of interest. On the other
hand, the factor $\sigma_0$ is a decreasing function of the smoothing
length. As a result, the genus asymmetry parameter is not a monotonic
function of the smoothing length. For example, the genus asymmetry
parameter in the $\Gamma = 0.2$, $\sigma_8 = 1$ CDM model is
approximately $0.08$ on wide range of smoothing length of $10\himpc
\simlt R \simlt 20\himpc$.

From the second order formula for the genus of equation
(\ref{eq2-207c}), the genus at the troughs is proportional to $-2 \pm
\sqrt{3} (S^{(2)} - S^{(0)})\sigma_0$. Therefore, the fraction of the
deviation from the Gaussian prediction at $\tnu = \pm \sqrt{3}$ is
$\pm \sqrt{3} (S^{(2)} - S^{(0)})\sigma_0/2$. When the genus asymmetry
parameter is $0.08$, this fraction is about $\pm 7\%$. Thus, the
perturbation theory predicts that the values of genus at positive and
negative troughs differ $14\%$ for the $\Gamma = 0.3$, $\sigma_8 = 1$
CDM model, and so on. This difference increases with the shape
parameter $\Gamma$. The degree of deviation is qualitatively
consistent with the numerical result in Fig.~9 of \citet{can98},
although quantitative comparison is still difficult because of the
noise in the simulation.

\begin{figure}
\epsscale{0.65} \plotone{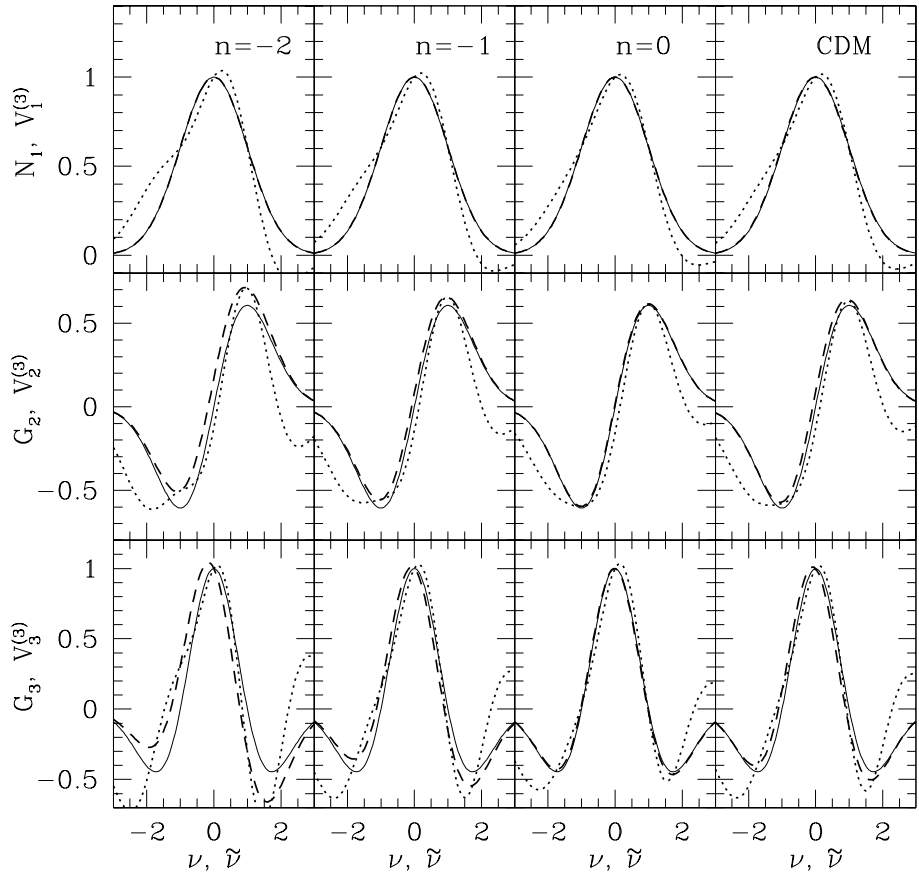}
\figcaption[f2.eps]{Same as Figure~\ref{fig1} for 3D velocity field.
\label{fig2}}
\end{figure}

The statistics for the velocity field is plotted in Figure~\ref{fig2}.
Since the the sign of the skewness parameters is negative in this
case, the weakly nonlinear evolution of statistics against the density
threshold $\nu$ indicates the sponge-like shift. In terms of the
volume-fraction threshold $\tnu$, on the other hand, meat-ball shifts
are observed even for relatively bluer spectrum with $\nspec \leq 0$.

\begin{figure}
\epsscale{0.4} \plotone{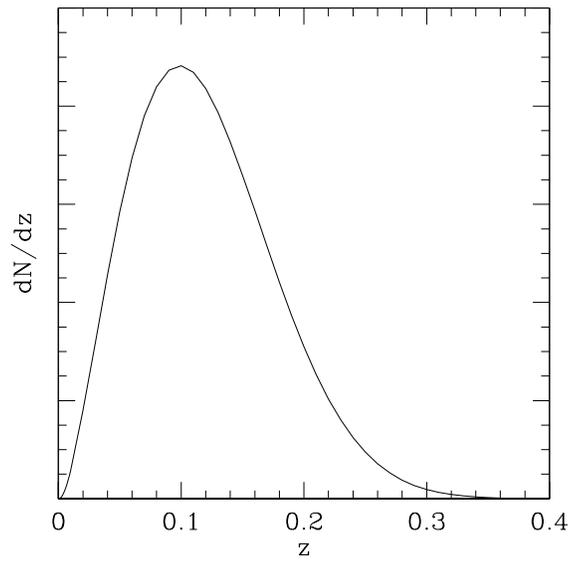} \figcaption[f3.eps]{The
differential number count for the APM luminosity function with B-band
limiting magnitude 19. The normalization is arbitrary.
\label{fig3}}
\end{figure}

The 2D projected galaxy statistics are dependent on the selection
function of galaxies and cosmological models. As an example, we assume
the APM luminosity function \citep{lav92} for galaxies with B band
magnitude limit $m_{\rm lim} = 19$. The differential number count
$dN/dz(z)$ for this sample is plotted in Figure~\ref{fig3}.
The resulting mean redshift is $\langle z \rangle = 0.12$. The
relation between the differential number count and the normalized mean
number density $n(\chi)$ in comoving coordinates are given by
\begin{eqnarray}
   n\left[\chi(z)\right] =
   \frac{\displaystyle H(z) \frac{dN}{dz}(z)}
      {\displaystyle \left\{S_K\left[\chi(z)\right]\right\}^2
         \int_0^\infty \frac{dN}{dz} dz},
\label{eq5-1}
\end{eqnarray}
where 
\begin{eqnarray}
&&
   H(z) = H_0
  \sqrt{(1 + z)^3 \Omega_0 + 
  (1 + z)^2 (1 - \Omega_0 - \lambda_0) + \lambda_0},
\label{eq5-2}\\
&&
   \chi(z) = \int_0^z \frac{dz'}{H(z')},
\label{eq5-3}
\end{eqnarray}
are the Hubble parameter and the comoving distance at redshift $z$,
respectively. Once the selection function $n(\chi)$ is given by
equation (\ref{eq5-1}), the integration of equations
(\ref{eq4-36})--(\ref{eq4-38}), and the interpolation of tabulated
values of $C^{(a)}$ in Table~\ref{tab6} give the skewness parameters.
For power-law power spectra, the skewness parameters are given by the
simpler integration of equation (\ref{4-50-2}) and the values of
$C^{(a)}$ in Table~\ref{tab7}.

\begin{figure}
\epsscale{0.65} \plotone{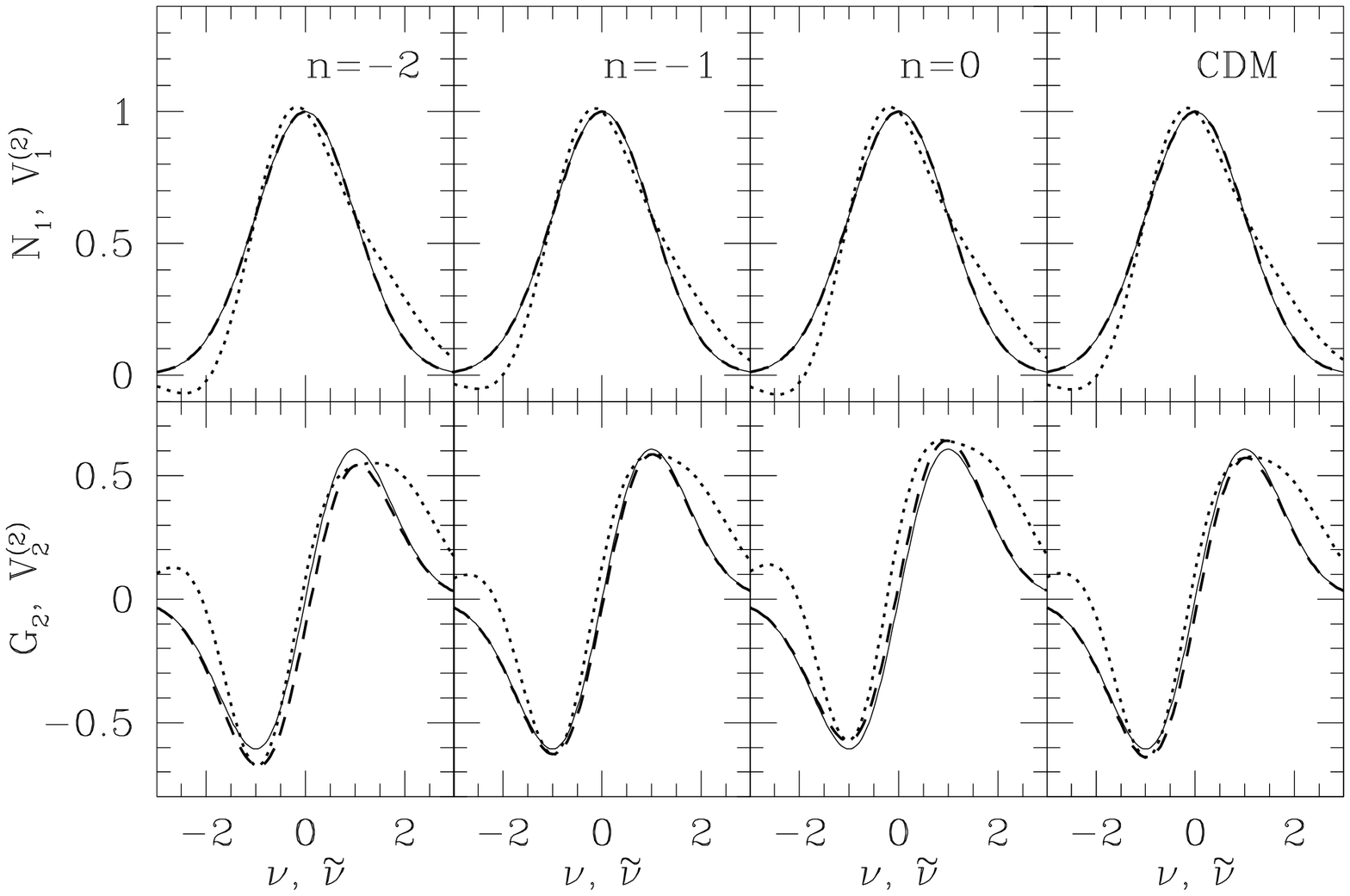} \figcaption[f4.eps]{The 2D genus
$G_2$, the level-crossing statistic $N_1$ and the Minkowski
functionals $V^{(3)}_k$ of the 2D projected density field. All curves
are appropriately normalized. The meaning of lines are the same as in
Figure~\ref{fig1}. The selection function from APM luminosity function
is assumed.
\label{fig4}}
\end{figure}

In Figure~\ref{fig4}, 2-dimensional statistics, $N_1$, $G_2$,
$V^{(2)}_1$ and $V^{(2)}_2$ are plotted with assumed cosmological
parameters, $\Omega_0=0.3$, $\lambda_0 = 0.7$, and APM luminosity
function with limiting magnitude 19.
The nonlinear parameter is assumed as $\sigma_0=0.2$. For the CDM
model, the shape parameter is assumed as $\Gamma = 0.25$, and we take
the smoothing angle as $\thf =1^{\circ}$. With this smoothing angle
the value $\sigma_0=0.2$ corresponds to the normalization $\sigma_8 =
1.22$. Basic features for these 2-dimensional statistics are the same
as for 3-dimensional density field, except that the smoothing scale in
CDM model we adopt corresponds to smaller scale than in the example of
3-dimensional density field of the Figure~\ref{fig1}.

\setcounter{equation}{0}
\section{CONCLUSIONS}
\label{sec6}

In this work, we comprehensively presented a basic formalism to treat
the statistics of smoothed cosmic fields in perturbation theory. This
formalism provides a methodology on evaluating how various statistics
deviate from the prediction of simple random Gaussian fields. As long
as the non-Gaussianity is weak, the behavior of the statistics caused
by non-Gaussianity is predicted by our formalism, which enable us to
quantitatively compare the statistical quantities and the source of
the non-Gaussianity. This method is considered as an extension of the
Edgeworth expansion, which has been proven to be useful in various
fields of research, as long as the non-Gaussianity is weak. In this
paper, we derive useful formulas and relations focusing on application
of the second-order perturbation theory to various cosmic fields.

Several examples of statistics of cosmic fields in second-order
perturbation theory are investigated in datail, including
level-crossing statistics, 2D and 3D genus statistics, 2D extrema
statistics, and the Minkowski functionals, which are extensively used
in cosmology. More complicated statistics, such as 2D and 3D density
peaks, can also be calculated, although they are more tedius.

A particular interest in cosmology of our method is in the application
to the cosmic fields. Even if the cosmic field was random Gaussian at
the initial stage, the gravitational evolution induces the
non-Gaussianity. The gravitational instability is a well-defined
process, so that we can evaluate the non-Gaussianity without any
ambiguity when the evolution remains in the quasi-linear regime
provided that the biasing from the non-gravitational process is simple
enough on large scales. Therefore, we performed the perturbative
analysis to obtain the necessary skewness parameters based on the
gravitational instability theory. We considered the 3D density field,
the 3D velocity field, the 2D projected density field. In the
application of second order theory to various statistics of smoothed
cosmic fields, three types of skewness parameters are commonly useful,
i.e., $S^{(0)}$, $S^{(1)}$, an $S^{(2)}$. Extensive calculations of
these parameters for various cosmic fields are one of the new results
of this paper. It would be true that other skewness parameters are
needed when other complex statistics are considered. Such other
parameters, if needed, are similarly calculated by the method we
outlined in this paper.

We find the lowest order deviations from the Gaussian predictions of
various statistics of smoothed cosmic fields depend only on the
differences of the skewness parameters when we use a threshold $\tnu$,
which is rescaled by a volume-fraction of the smoothed field. This
rescaling makes the lowest deviations much smaller than in the case of
direct threshold $\nu$. This is because three skewness parameters
$S^{(a)}$ take similar values if it is arisen from the gravitational
evolution. For the phenomenological hierarchical model, these
parameters are identical. In this case, the weakly nonlinear
correction of the statistical quantities in terms of the
volume-fraction threshold vanishes. When evaluated by the second-order
perturbation theory of density fluctuations, those three types of
skewness parameter are still close to each other. This fact explains
the smallness of the deviations from Gaussian predictions of
statistical quantities like genus, level-crossing, or Minkowski
functionals when the rescaled threshold by volume-fraction is used.

We discussed small, but detectable deviations from Gaussian prediction
of the 3D genus curve against rescaled threshold in detail. In the
framework of the second-order perturbation theory, a prominent
deviation of the genus curve occurs at the two troughs of the curve.
Relative depths of left and right troughs in the genus curve show
slightly meat-ball shift for CDM-like models. We found the degree of
this asymmetry is proportional to the combination $(S^{(2)} -
S^{(0)})\sigma_0$. We call this factor as a genus asymmetry parameter,
which we propose as theory-motivated parameter that characterize the
asymmetry of the genus curve. The genus asymmetry parameter can
observationally be obtained by fitting the genus curve by Hermite
polynomials as equation (\ref{eq5-0}). Qualitative comparison with the
numerical simulations in literatures suggests that such asymmetry can
actually be observable. We have not estimated over what dynamic range
in nonlinearity parameter like $\sigma_0$, $\nu\sigma_0$, or $\tnu
\sigma_0$ are each perturbative expressions valid. This estimation
requires a systematic comparison with large N-body simulations, which
is beyond the scope of this paper, and will be given in a subsequent
paper of the series.

In principle, any order in the perturbation theory can be calculated
as further as one would like. Although the computation of the
higher-order theory becomes more and more tedious, the necessity of
the comparison with large-scale cosmological observations is a good
reason to perform such computation as further as we can. One of the
spectacular example of the detailed comparison between perturbation
theory and observations is the fine structure constant in quantum
electrodynamics \citep[e.g.,][]{kin96}. Our analysis in this paper
will be extended to the third-order perturbation theory in a
subsequent paper of the series. The present time is in an unique
decade when the observations of cosmic fields are in unforeseen
progress, like large-scale redshift surveys, detailed mapping of CMB
fluctuations, gravitational lensing surveys, and so forth. Statistics
of smoothed density field with higher-order perturbation theory will
provide an unique method to analyze those high-precision data. The
precision cosmology is undoubtedly providing clues to unlock the door
to the origin of the universe.

\acknowledgments

I would like to thank M.~Kerscher and B.~Jain for discussions. I wish
to acknowledge support from JSPS Postdoctoral Fellowships for Research
Abroad, and from the Ministry of Education, Culture, Sports, Science,
and Technology, Grant-in-Aid for Encouragement of Young Scientists,
13740150, 2001.

\appendix

\section{USEFUL GAUSSIAN INTEGRALS}
\label{app1}
\setcounter{equation}{0}

In this appendix, we give Gaussian integrals which are useful in this
paper. In the following, $H_n$ is the Hermite polynomials,
\begin{eqnarray}
   H_n(\nu) =
   e^{\nu^2/2}
   \left(- \frac{\partial}{\partial\nu} \right)^n
   e^{-\nu^2/2},
\label{eqa1-1}
\end{eqnarray}
and we further employ the notation
\begin{eqnarray}
   H_{-1}(\nu) \equiv
   e^{\nu^2/2}
   \int_\nu^\infty d\nu
   e^{-\nu^2/2} =
   \sqrt{\frac{\pi}{2}} e^{\nu^2/2}
   {\rm erfc}\left(\frac{\nu}{\sqrt{2}}\right).
\label{eqa1-2}
\end{eqnarray}
Several Hermite polynomials are
\begin{eqnarray}
   &&
   H_0(\nu) = 1, \quad
   H_1(\nu) = \nu, \quad
   H_2(\nu) = \nu^2 - 1, \quad
   H_3(\nu) = \nu^3 - 3\nu,
\nonumber\\
   &&
   H_4(\nu) = \nu^4 - 6\nu^2 + 3, \quad
   H_5(\nu) = \nu^5 - 10\nu^3 + 15\nu.
\label{eqa1-2-1}
\end{eqnarray}
The Hermite polynomial at zero is given by
\begin{eqnarray}
   H_n(0)
   &=&
   \left\{
   \begin{array}{ll}
      0, & (n: {\rm odd}) \\
      \displaystyle
      (-1)^{n/2} (n-1)!! & (l: {\rm even})
   \end{array}
   \right.
\nonumber\\
   &\equiv& h_n.
\label{eqa1-3}
\end{eqnarray}
We generalize the above definition of $h_n$ to the case $n<0$ by
interpreting $(n-1)!!$ as the appropriate gamma function so that
$(-1)!! = 1$, $(-3)!! = -1$, etc. For example, $h_{-2} = 1$, $h_{0} =
1$, $h_{2} = -1$, $h_{4} = 3$, and so forth. In this appendix, we give
useful Gaussian averages $\langle\cdots\rangle_{\rm G}$ of equation
(\ref{eq1-12-1}) for normalized cosmic field $\alpha$ defined by
equation (\ref{eq1-3}) and its spatial derivatives $\eta_i =
\alpha_{,i}$. In the following, $\eta$ represents any one of the
components $\eta_i$. First, concerning $\alpha$,
\begin{eqnarray} &&
   \left\langle
      \frac{\partial^k\delta(\alpha - \nu)}{\partial\alpha^k}
      H_n(\alpha)
   \right\rangle_{\rm G} = 
   \frac{e^{-\nu^2/2}}{\sqrt{2\pi}} H_{k+n}(\nu),
\label{eqa1-4a}\\
&&
   \left\langle
      \frac{\partial^k \theta(\alpha - \nu)}{\partial\alpha^k}
      H_n(\alpha)
   \right\rangle_{\rm G} = 
   \frac{e^{-\nu^2/2}}{\sqrt{2\pi}} H_{k+n-1}(\nu).
\label{eqa1-4b}\\
\end{eqnarray}
Second, concerning $\eta$, in the notation of equation (\ref{eqa1-3}),
\begin{eqnarray}
&&
   \left\langle
      \frac{\partial^l |\eta|}{\partial\eta^l}
   \right\rangle_{\rm G} = 
   \sqrt{\frac{2}{\pi}}
   \left(
      \frac{\sigma_1}{\sqrt{d}\sigma_0}
   \right)^{1-l} h_{l-2},
\label{eqa1-5a}\\
&&
   \left\langle
      \frac{\partial^l \delta(\eta)}{\partial\eta^l}
   \right\rangle_{\rm G} = 
   \sqrt{\frac{1}{2\pi}}
   \left(
      \frac{\sigma_1}{\sqrt{d}\sigma_0}
   \right)^{-l-1} h_l.
\label{eqa1-5b}
\end{eqnarray}

\section{LIMBER'S EQUATION FOR BISPECTRUM}
\label{app2}
\setcounter{equation}{0}

The correlation functions on projected sky is expressible by the
3-dimensional correlation functions. The explicit relation for the
two-point correlation function is given by Limber's equation
\citep{lim54}. The Fourier-space version of Limber's equation is given
by Kaiser \citep{kai98} and is somewhat simpler. His argument was
generalized to higher-order correlation functions and their Fourier
transforms \citep{sco99,buc00}. Since the higher-order Limber's
equation in Fourier space was discussed in the context of weak lensing
field in literatures, here we review the derivation in a way more
useful in this paper. Let 2D projected field $f$ be the projection of
a time-dependent 3D field $F(\bfx;\tau)$ along a light-cone:
\begin{eqnarray}
   f(\bfth) = 
   \int d\chi S_K^{\,2}(\chi) q(\chi) 
   F(\chi, \bfth S_K(\chi); \tau_0 - \chi), \label{eqa2-1}
\end{eqnarray}
where $q(\chi)$ is some radial weighting function and $\chi$ is the
radial comoving distance, and $\tau_0$ is the conformal time at the
observer. The past light-cone of the observer is specified by the
equation $\chi = \tau_0 - \tau$. The comoving angular distance
$S_K(\chi)$ is defined by equation (\ref{eq4-29}).

In the following, we explicitly derive the relation for 3-point
correlation function, and bispectrum. From the Limber's equation,
power spectrum is already derived by \citet{kai98}:
\begin{eqnarray} &&
   P_f(\omega) =
   \int d\chi S_K^{\,2}(\chi) q^2(\chi)
   P_F(\frac{\omega}{S_K(\chi)};\tau_0-\chi),
\label{eqa2-2}
\end{eqnarray}
where $P_f$ and $P_F$ is the power spectrum of fields $f$ and $F$,
respectively. We generalize this equation to the one for 3-point
statistics below. The generalization of the following derivation to
higher-order statistics is straightforward. The angular 3-point
correlation function $w^{(3)}_f$ of $f$ is
\begin{eqnarray}
   w^{(3)}_f(\bfth_1, \bfth_2, \bfth_3) &=&
   \int d\chi_1 S_K^{\,2}(\chi_1) q(\chi_1)
   \int d\chi_2 S_K^{\,2}(\chi_2) q(\chi_2)
   \int d\chi_3 S_K^{\,2}(\chi_3) q(\chi_3)
\nonumber\\
&& \qquad \times
   \left\langle 
      F\left(
         \chi_1, \bfth_1 S_K\left(\chi_1\right); \tau_0 - \chi
      \right)
      F\left(
         \chi_2, \bfth_2 S_K\left(\chi_2\right); \tau_0 - \chi
      \right)
      F\left(
         \chi_3, \bfth_3 S_K\left(\chi_3\right); \tau_0 - \chi
      \right)
   \right\rangle
\nonumber\\
   &\simeq&
   \int d\chi {S_K}^6(\chi) q^3(\chi)
   \int d\chi_1 d\chi_{2}
   \zeta_F\left(
      \chi_1, \bfth_1 S_K(\chi_1);\chi_2, \bfth_2 S_K(\chi_2);
      \chi, \bfth_3 S_K(\chi);\tau_0 - \chi
   \right),
\label{eqa2-3}
\end{eqnarray}
where $\zeta_F(\bfx_1;\cdots;\bfx_3;\tau)$ is the spatial 3-point
correlation function of the field $F$ with the 3-point configuration
$(\bfx_1;\bfx_2;\bfx_3)$ at conformal time $\tau$. According to the
spirit of the Limber's equation, we assume that $S_K^{\,2}(\chi)
q(\chi)$ is slowly varying compared to the scale of the fluctuations
of interest and also that these fluctuations occur on a scale much
smaller than the curvature scale. The equation (\ref{eqa2-3}) is the
generalization of the Limber's equation to higher-order correlation
functions.

Now we transform equation (\ref{eq2-13}) to obtain the 2D bispectrum.
We use the following convention of the Fourier transforms
\begin{eqnarray}
&&
   \widetilde{f}(\bfom) =
   \int d^2\theta f(\bfth) e^{-i\sbom\cdot\sbth},
\label{eqa2-4a}\\
&&
   \widetilde{F}(\bfk;\tau) =
   \int d^3x F(\bfx;\tau) e^{-i\sbk\cdot\sbx},
\label{eqa2-4b}
\end{eqnarray}
the bispectrum $B_f$ of the 2D field $f$, and $B_F$ of the 3D field
$F$ are defined by
\begin{eqnarray}
&&
   \left\langle 
      \widetilde{f}(\bfom_1) \widetilde{f}(\bfom_2) 
      \widetilde{f}(\bfom_3)
   \right\rangle =
   (2\pi)^2 \delta^2(\bfom_1 + \bfom_2 + \bfom_3)
   B_f(\omega_1,\omega_2,\omega_3),
\label{eqa2-5a}\\
&&
   \left\langle 
      \widetilde{F}(\bfk_1;\tau)\widetilde{F}(\bfk_2;\tau)
      \widetilde{F}(\bfk_3;\tau)
   \right\rangle =
   (2\pi)^3 \delta^3(\bfk_1 + \bfk_2 + \bfk_3)
   B_F(k_1, k_2, k_3;\tau),
\label{eqa2-5b}
\end{eqnarray}
where $\omega_i = |\bfom_i|$ and $k_i = |\bfk_i|$. The Dirac's delta
function comes from the translational invariance of statistics. From
the relations,
\begin{eqnarray}
&&
   B_f(\omega_1,\omega_2,\omega_3)
   =
   \int d^2\theta_1 d^2\theta_2
   v_f(\bfth_1,\bfth_2,{\bf 0})
   e^{- i\sbom_1\cdot\sbth_1 - i\sbom_2\cdot\sbth_2},
   \quad (\omega_3 = |\bfom_1 + \bfom_2|),
\label{eqa2-6a}\\
&&
   \zeta_F\left(\bfx_1,\bfx_2,\bfx_3;\tau\right)
   =
   \int \frac{d^3k_1}{(2\pi)^3} \frac{d^3k_2}{(2\pi)^3}
   B_F(k_1,k_2,k_3;\tau)
   e^{i\sbk_1\cdot(\sbx_1-\sbx_3) + i\sbk_2\cdot(\sbx_2-\sbx_3)},
\label{eqa2-16b}
\end{eqnarray}
the Fourier transform of the equation (\ref{eqa2-3}) reduces to a
simple equation,
\begin{eqnarray}
   B_f(\omega_1,\omega_2,\omega_3) =
   \int d\chi S_K^{\,2}(\chi) q^3(\chi)
   B_F(\frac{\omega_1}{S_K(\chi)},\frac{\omega_2}{S_K(\chi)},
      \frac{\omega_2}{S_K(\chi)};\tau_0-\chi).
\label{eqa2-7}
\end{eqnarray}
This is a bispectrum version of the Limber's equation and the
generalization of the Kaiser's equation for power spectrum
(\ref{eqa2-2}).

\section{Symbol Index}
\label{app3}
\setcounter{equation}{0}

In Table~\ref{tab9}, the quantites used in this paper are listed.
\begin{deluxetable}{ccc}
\tablecolumns{3}
\tablewidth{0pt}
\tablecaption{Symbol Index
\label{tab9}}
\tablehead{
\colhead{Symbol} & \colhead{Definition} & \colhead{Equation}}
\startdata
$\langle\cdots\rangle$ & Ensemble average & (\ref{eq1-12a}) \\
$\langle\cdots\rangle_{\rm G}$ & Ensemble average by Gaussian field &
(\ref{eq1-12-1}) \\ 
$\Sigma_j$  & 2D spectral parameter & (\ref{eq4-36})\\ 
$\Omega$ & Time-dependent density parameter & (\ref{eq4-34-1})\\
$\Omega_0$ & Density parameter & (\ref{eq4-34-1})\\
$\alpha$ & Normalized field of $f$ & (\ref{eq1-3})\\
$\alpha_{,ij}$, etc. & $\partial^2 \alpha/\partial x_i \partial x_j =
\partial_i \partial_j \alpha$, etc. &
(\ref{eq1-4})\\ 
$\gamma$ & spectral parameter & (\ref{eq1-13-5})\\
$\delta$ & Mass density contrast & (\ref{eq4-70})\\
$\delta_{\rm 3D}$ & 3D density contrast & (\ref{eq4-31})\\
$\delta_{\rm g}$ & Galaxy number density contrast & (\ref{eq4-70})\\
$\zeta_{ij}$ & $=\alpha_{,ij}$, Tensor notation & {}\\
$\widetilde{\zeta}_{ij}$ & Linear combination of $\alpha$ and
$\zeta_{ij}$ & (\ref{eq1-13-3-1})\\ 
$\eta_i$ & $= \alpha_{,i}$, Vector notation & {}\\
$\bfth$ & Angular coordinates in a small patch of the sky & (\ref{eq4-31})\\
$\kappa$ & Local convergence field & (\ref{eq4-60})\\
$\lambda$ & Time-dependent scaled cosmological constant & (\ref{eq4-34-1})\\
$\lambda_0$ & Scaled cosmological constant & (\ref{eq4-34-1})\\
$\nu$ & Threshold by variance & {}\\
$\widetilde{\nu}$ & Threshold by volume fraction & (\ref{eq2-105})\\
$\rho$ & 3D comoving density field & (\ref{eq4-30})\\
$\rho_{\rm e2}$ & 2D Weighted extrema & (\ref{eq2-24})\\
$\rho_{\rm p}$ & 2D projected density field & (\ref{eq4-30})\\
$\sigma_0$ & Variance of $f$ & (\ref{eq1-2})\\
$\sigma_1$, $\sigma_2$  & Spectral parameters &
(\ref{eq1-13-3a}), (\ref{eq1-13-3b})\\ 
$\tau_0$ & Conformal time at the present & (\ref{eq4-30})\\
$\chi$ & Comoving distance & (\ref{eq4-29})\\ 
$\omega_k$ & Volume of the unit ball in $k$ dimensions &
(\ref{eq2-54})\\ 
$A_\mu, \bfA$ & Vector of spatial derivatives of $\alpha$ & (\ref{eq1-4})\\
$B(k_1,k_2,k_3)$ & Bispectrum at the present time & (\ref{eq4-4b})\\
$B_{\rm 2D}(\omega_1,\omega_2,\omega_3)$ & 2D bispectrum & (\ref{eq4-33b})\\
$B_{\rm 3D}(k_1,k_2,k_3;\tau)$ & 3D bispectrum at conformal time $\tau$ &
(\ref{eq4-33b})\\ 
${C}^{(a)}$ & Integrand of the 2D skewness parameters &
(\ref{eq4-39})\\
$\widetilde{C}^{(a)}$ & Integrand of the 2D skewness parameters &
(\ref{eq4-40})\\
$C^{\alpha\beta}_m$ & Contributions to the 2D skewness parameters &
(\ref{eq4-46})\\
$D$ & linear growth rate & (\ref{eq4-34a})\\
$E$ & Parameter of the perturbation theory & (\ref{eq2-21-1})\\
$E_v$ & Parameter of the perturbation theory & (\ref{eq4-24b})\\
$F$ & Cosmic field, in general & (\ref{eq1-12a})\\
$F_{,\mu_1\mu_2\mu_3}$, etc. & 
$\partial^3 F/\partial A_{\mu_1} \partial A_{\mu_2}\partial A_{\mu_3}$,
etc. & (\ref{eq1-13}) \\
$F(\alpha,\beta,\gamma;z)$ & Gauss hypergeometric function ${}_2 F_1$ &
(\ref{eq4-20})\\ 
$G_2$ & 2D genus statistic & (\ref{eq2-8}), (\ref{eq2-9})\\
$G_3$ & 3D genus statistic & (\ref{eq2-15}), (\ref{eq2-17})\\
$H$ & Time-dependent Hubble parameter & (\ref{eq4-23})\\
$H_0$ & Hubble constant & (\ref{eq5-2})\\
$I_\nu$ & Modified Bessel function & (\ref{eq4-11})\\
$J_\mu, \bfJ$ & Fourier counterpart of $A_\mu$ & (\ref{eq1-5})\\
$J^{(2)}_m(\{m_{ij}\})$ & Integers defined by Table~\ref{tab1} &
(\ref{eq2-19})\\
$K$ & Spatial curvature & (\ref{eq4-29})\\
$H_n(\nu)$ & Hermite polynomials & (\ref{eqa1-1})\\
$H_{-1}(\nu)$ & Extended Hermite polynomial of order $-1$ & (\ref{eqa1-2})\\
$M^{(n)}_{\mu_1 \cdots \mu_n}$ & $n$-th order cumulants of $A_\mu$ &
(\ref{eq1-6})--(\ref{eq1-7d})\\
$\widehat{M}^{(n)}_{\mu_1 \cdots \mu_n}$ & Normalized cumulants of $A_\mu$ &
(\ref{eq1-12})\\
$\bfM$ & second-order cumulant of $A_\mu$ & (\ref{eq1-8})\\
$N_1$ & Level-crossing statistic & (\ref{eq2-1a})\\
$N_2$ & Length statistic & (\ref{eq2-1b})\\
$N_3$ & Area statistic & (\ref{eq2-1c})\\
$P_{\rm G}$ & Multivariate Gaussian distribution function & (\ref{eq1-5})\\
$P$ & Multivariate probablitiy distribution function & (\ref{eq1-5})\\
$P(k)$ & Power spectrum at present & (\ref{eq4-4a})\\
$P_{\rm 2D}(\omega)$ & 2D power spectrum & (\ref{eq4-33a})\\
$P_{\rm 3D}(k;\tau)$ & 3D power spectrum at conformal time $\tau$ &
(\ref{eq4-33a})\\ 
$\PL(k)$ & Linear power spectrum & (\ref{eq4-7a})\\
$Q$ & Hierarchical amplitude of the 3-point function &
(\ref{eq4-1})\\ 
$S^{(a)}$, $S_2^{(2)}$ & Skewness parameters &
(\ref{eq1-15a})--(\ref{eq1-15d})\\
$\widetilde{S}^{(a)}$ & Integrand of the skewness parameters &
(\ref{eq4-9})\\
$S^{\alpha\beta}_m(R)$ & Contributions to the skewness parameters &
(\ref{eq4-13})\\
$S_{\rm g}^{(a)}$ & Galaxy skewness parameters & (\ref{eq4-29})\\
$S_K$ & Comoving angular diameter distance & (\ref{eq4-29})\\
$T_{\rm CDM}$ & CDM transfer function & (\ref{eq4-15})\\
$V^{(d)}_k$ & Minkowski functionals & (\ref{eq2-50-0}),
(\ref{eq2-50}), (\ref{eq2-52}), (\ref{eq2-54}) \\
$W$ & Smoothing function in real space & (\ref{eq1-1})\\
$W_R$ & Smoothing function in real space & (\ref{eq1-1})\\
$Z$ & Generating function & (\ref{eq1-5})\\
$b$ & Bias parameter & (\ref{eq4-70})\\
$b_2$ & Nonlinear bias parameter & (\ref{eq4-70})\\
$d$ & Dimension of the sample space & (\ref{eq1-1})\\
$dN/dz$ & Differential number count & (\ref{eq5-1})\\
$f$ & Smoothed cosmic field, in general & (\ref{eq1-1})\\
$f_V$ & Volume fraction on the high-density side & (\ref{eq2-105})\\
$\logd$ & Logarithmic derivative of the growth factor & (\ref{eq4-25})\\
$h_l$ & Hermite polynomial at zero, $H_l(0)$, extended to $l < 0$ &
(\ref{eqa1-3})\\ 
$n$ & Selection function per unit comoving volume & (\ref{eq4-30})\\
$\nspec$ &index for the power-law power spectrum & (\ref{eq4-16})\\
$x$, $y$, $z$ & Linear combinations of $\alpha$ and $\zeta_{ij}$'s &
(\ref{eq1-13-4a})--(\ref{eq1-13-4c})\\ 
\enddata
\end{deluxetable}

\clearpage


\end{document}